\documentclass[aps,prb,twocolumn,groupedaddress,superscriptaddress,nobibnotes]{revtex4-2}
\usepackage{amsmath}
\usepackage{amssymb}
\usepackage{graphicx}
\usepackage{dcolumn}
\usepackage{bm}
\usepackage{xcolor}
\usepackage[yyyymmdd,hhmmss]{datetime}
\usepackage[normalem]{ulem}

\usepackage{siunitx}
\usepackage{soul}
\setulcolor{red}
\usepackage[linkcolor=blue,urlcolor=blue,citecolor=blue,colorlinks=true]{hyperref}

\begin{document}

\title{Unveiling the superconducting scenario in multiphase superconductor CeRh$_2$As$_2$ from space-group symmetry analysis and DFT calculations}

\author{V.\,G. Yarzhemsky}
\affiliation{Saint Petersburg State University, Saint Petersburg,
Russia, 199034} \affiliation{Kurnakov Institute of General and
Inorganic Chemistry of RAS, 119071, Moscow, Russia}

\author{E.\,A. Teplyakov}
\affiliation{Saint Petersburg State University, Saint Petersburg,
Russia, 199034} \affiliation{Kurnakov Institute of General and
Inorganic Chemistry of RAS, 119071, Moscow, Russia}

\author{S.\,V. Eremeev}
\affiliation{Institute of Strength Physics and Materials Science of
RAS,  634055 Tomsk, Russia} \affiliation{Saint Petersburg State
University, Saint Petersburg, Russia, 199034}

\author{E.V.~Chulkov}
\affiliation{Donostia International Physics Center, 20018
Donostia-San Sebasti\'an, Spain} \affiliation{Saint Petersburg State
University, Saint Petersburg, Russia, 199034}

\date{\today}

\begin{abstract}
Despite of the low transition temperature, the recently identified superconductor CeRh$_2$As$_2$ has garnered significant interest due to its unique symmetry and magnetic characteristics, particularly the existence of two superconducting (SC) phases under a magnetic field, one of which exceeds the Pauli-Clogston limit. The field-induced transition from a low-field even-parity state to a high-field odd-parity state is usually described as a singlet-triplet transition. However, it is uncommon for a single compound to exhibit both triplet and singlet SC scenarios. The aim of this paper is to investigate the possibilities of symmetry changes in the SC state without a change of spin multiplicity. To this end, we construct the SC order parameter based on Anderson pair functions, considering the phase winding within the symmetry of the point group $D_{4h}$ and the magnetic group $4/mm^{\prime }m^{\prime}$. It was found that two triplets with opposite-spin and equal-spin pairing states of symmetry $E_{1u}^{\prime +}$, are nodeless but exhibit distinct internal structures and may be associated with low-and high-field phases. Additionally, nontrivial Cooper pairing resulting from the non-symmorphic structure of the space group was examined, particularly in the case where the Fermi surface intersects with the boundaries of a Brillouin zone (BZ). It was determined that at the X point, triplet pairs are even, while singlet pairs can be either even or odd. Furthermore, at the X point, pair density waves that alter phase by $\pi$ at the atomic centers linked by lattice translations are also feasible. To explore the possibility of such scenarios, precise DFT calculations of the band structure were performed, revealing the contribution of Ce $4f$ electrons to the states at the Fermi level. Thus, the even-odd transition can take place in a triplet scenario at symmetry points of a BZ.
\end{abstract}

\maketitle

\section{Introduction}

In recently discovered superconductor CeRh$_{2}$As$_{2}$, the
superconducting critical field of its high-field phase is as high as
14 T, although the transition temperature $T_{c}$ is only 0.26 K
\cite{khim1}. Additionally, there is some instability in the state
at $T_0= 0.4$ K. When the magnetic field is applied along the
crystallographic $c$ axis of the tetragonal cell, the
superconducting state experiences a phase transition at $H^\ast \sim
4$ T, which was identified by several thermodynamic probes
\cite{khim1}. The superconducting transition is completely
suppressed down to 0.05 K at magnetic fields of 14 T for $H\| c$ and
2 T for $H\| ab$. The upper critical field of the high-field phase
denoted, as SC2, is not Pauli-Clogston-paramagnetically suppressed.
In contrast, the low-field superconducting state (SC1) is strongly
Pauli-Clogston-limited \cite{khim1}. The phases SC1 and SC2 were
interpreted as even and odd, respectively \cite{khim1}. Remarkably,
the SC phase-switching field $H^{\ast}$ is drastically reduced under
pressure, dropping to 0.3 T at 2.7 GPa \cite{Semeniuk_PRB2024}. It
was suggested that the key superconducting properties of
CeRh$_2$As$_2$ are likely connected with the the absence of local
inversion symmetry in an overall inversion-symmetric crystal
structure, $P4/nmm$ (No. 129, $D^7_{4h}$) \cite{khim1}.

The field-angle dependence of the upper critical field, as
determined through magnetic $ac$ susceptibility and specific heat
measurements on single crystals of CeRh$_2$As$_2$  has demonstrated
the presence of odd-parity superconductivity in the SC2 phase, which
is consistent with the $d$-vector in the $ab$ plane
\cite{Landaeta_PRX2022}. Note that $d$-vector in the $ab$ plane
corresponds to a triplet spin projection  $S_{z}=\pm 1$, called
equal spin pairing (ESP), and the $d$-vector normal to the plane
corresponds to $S_{z}=0$, called the opposite spin pairing (OSP)
case \cite{maeno12}.   Earlier DFT calculations indicated that the
Fermi surface is primarily formed by the Ce $4f$ electrons
hybridizing with the Rh $4d_{x^2-y^2}$ electrons, with the exception
of the pockets near the $\mathrm{A}$ point \cite{nogaki}.  Drawing
from the similarities between experimental results and theoretical
predictions  \cite{yoshida}, a  pair density wave (PDW) scenario has
been proposed for CeRh$_2$As$_2$.

In order to explain the existence of even and odd pairs in a single
material, several models were proposed. The model in which the
dominant interaction in pairs of electrons on the same sublattice is
a spin singlet and the sublattices are swapped by space inversion
$I$,  resulting in the formation of even and odd-parity states, was
suggested \cite{Cavanagh_PRB2022}. It has also been proposed that
CeRh$_{2}$As$_{2}$ hosts a quadrupole density wave (QDW) instability
and the anomaly at $T_0$ is a signature of the quadrupolar degrees
of freedom \cite{Hafner_PRX2022}. The $^{75}$As  nuclear magnetic
resonance (NMR) measurements revealed the presence of the
antiferromagnetic (AFM) order  within the SC1 phase and indicated
that the AFM state disappears at the transition to the high-field
SC2 phase \cite{Ogata_PRB2024}. Moreover, it was reported that the
AFM order is inside the SC1 phase and that the    AFM state
disappears at the transition field to the high-field SC2 phase
\cite{Ogata_PRL2023}.

The transition between SC1 and SC2 in CeRh$_2$As$_2$ was   regarded
as an    even-odd parity within the $C_{4h}$ symmetry and  the
superconducting order parameter (SOP) was expressed using linear
combinations of spherical functions \cite{nogaki2,
Ishizuka_PRB2024}. Such a representation of the SOP resulted in the
presence of solely point nodes of $A_{u}$  in the  $k_z$ direction
in $C_{4h}$ symmetry \cite{Ishizuka_PRB2024}.

Triplet pairing mechanisms derived from Hund's--Kondo model were
also developed for CeRh$_2$As$_2$ \cite{Hazra_PRL2023}, and the role
of non-symmorphic symmetry was pointed out.  It is important to note
that Hund's-like models assume one-center triplet correlations
\cite{Hotta_PRL2004}. However,  specific heat measurement outcomes
characterize CeRh$_2$As$_2$  as an antiferromagnetic superconductor,
which  undergoes a metamagnetic-like phase transition at $T_0$ when
subjected to magnetic fields applied along the $c$ axis. This
transition is associated with alterations in the spin structure and
another change in the magnetic structure is observed at $H\approx
4$~T. Neutron scattering experiments have identified AFM spin
correlations,   indicating that superconductivity in CeRh$_2$As$_2$
is mediated by AFM spin fluctuations \cite{Chen_PRL2024}. The muon
spin relaxation ($\mu$SR) studies have demonstrated the coexistence
of local magnetism and superconductivity  \cite{Khim_PRB2025}.
Recent magnetic field-dependent measurements of magnetostriction and
$ac$ susceptibility confirmed the coexistence of the AFM phase that
appears in the $H\|c$ field at $T_0$ within both SC1 and SC2 phases
\cite{khanenko2025}. The results of local magnetization measurements
suggest fully-gapped superconductivity in the SC1 phase; however,
the possibility of two-band superconductivity with a small $d$-wave
gap function has not been ruled out \cite{juraszek2025}.

Photoelectron spectroscopy measurements showed signs of  Ce-$4f$
nesting on the Fermi surface at the X point \cite{Wu_CPL2024}. Since
at points on Brillouin zone (BZ) faces direct connection between
multiplicity and parity of a Cooper pair may be violated
\cite{Yarzhemsky1992} the pairing symmetry  at this points may
differ from that at generic  points of a BZ. Therefore, in this
work, special attention in the group-theoretical consideration and
in the DFT calculations was given to such points.

A transition changing symmetry between two SC phases of
CeRh$_{2}$As$_{2}$  is typically regarded as a singlet-triplet
transition. Nevertheless, the spin multiplicity determines the
permutation symmetry, which in turn defines the sign of an exchange
matrix element.  Given the underlying magnetic structures, the
possibility of Cooper pairing resulting from the exchange
interaction between Ce $4f$ electrons is considered the most likely
scenario. Consequently, the presence of phases with different
multiplicity would imply the presence of two different scenarios of
superconductivity in a single material, which is unlikely.

The aim of this work is to study the possibility of changing the
symmetry of Cooper pairs without changing the spin multiplicity of
the pairs. It is worth noting that for a generic point in a BZ, the
spatial part of a singlet (triplet) pair is even (odd) with respect
to space inversion, as demonstrated in both the phenomenological
\cite{Sigrist_RMP1991} and space-group \cite{Yarzhemsky1992}
approaches. However, this direct relationship may be violated   at
BZ faces \cite{Yarzhemsky1992,Micklitz_PRL2017,Nomoto2017}. Earlier
DFT calculations and photoelectron spectroscopy measurements have
revealed multiple intersections of the Fermi surface (FS) with the
BZ faces, attributed to the contributions of Ce $4f$ electrons.
Consequently, it can be inferred that the band structure of
CeRh$_{2}$As$_{2}$ is favorable to the appearance of Cooper pairs
with unconventional symmetry.

In the present work SOP in phases SC1 and SC2 is considered making
use of the space-group approach to the wavefunction of Cooper pair
\cite{Yarzhemsky1992,Yarzhemsky1998}. This approach has previously
yielded significant insights  for non-symmorphic groups on the BZ
faces  \cite{Micklitz_PRL2017,Nomoto2017}. To support our findings,
the DFT calculations for CeRh$_{2}$As$_{2}$ were performed to
confirm the Ce-$4f$ contribution at the intersections of FS with the
BZ boundaries. Our analysis revealed intriguing relationships
between the multiplicity and parity of the pair function at
high-symmetry points M, X, and Z  as well as on the boundary of  BZ.
Specifically, we identified possible symmetries of PDW  at the X
point. For the FS inside  BZ the SOP was constructed on the basis of
Anderson functions \cite{Anderson_PRB1984} incorporating phase
winding for the point group $D_{4h}$ and magnetic group
$4/mm^{\prime}m^{\prime}$ \cite{Teplyakov2025,Yarzhemsky2021}.
Additionally, we considered the spin-orbit coupling as direct
products of   spatial and spin parts of the Cooper pair wavefunction
in point and magnetic group symmetry. Application of this technique
revealed the possibility of a transition SC1 $\to$ SC2  with a
change in the structure of the pair, but without  altering its
multiplicity, namely as an OSP $\to$ ESP  transition. To  explore
this possibility we constructed the SOP  from functions of
individual Cooper pairs, taking into account  both    spin-orbital
coupling and phase winding \cite{Teplyakov2025,Yarzhemsky2021}.

\section{Methods}

\subsection{ Building of pair function and SOP}

\subsubsection{ Spatial part of pair function }

\begin{table*}[ht!]
\caption{\label{tab:table1} Possible  symmetries of  spatial parts
of singlet and triplet pairs for the one-electron wavevector $k$ at
symmetry points, direction and planes in a BZ for space group
$P4/nmm$ ($D_{4h}^{7}$); $t_{i}$ are projective representations
\cite{kov}. }
    \centering
    \begin{ruledtabular}
\begin{tabular}{|l|l|l|l|l|}
$k$ & IR($H$) & singlet pair & triplet pair & $d_{\alpha }$  \\
 \hline
planes &  &  &   &  \\
 \hline
$(010),k_{y}=0$ & all & $\ A_{1g}+B_{1g}+E_{g}$ & $A_{2u}+B_{2u}+E_{u}$ & $I$ \\
$(010),k_{y}=\pi $ &all & $A_{2g}+B_{2g}+E_{g}$ & $A_{1u}+B_{1u}+E_{u}$ & $I$\\
(001)& all & $A_{1g}+A_{2g}+B_{1g}+B_{2g}$ & $2E_{u}$  & $I$\\
$(110)$ & all & $\ A_{1g}+B_{2g}+E_{g}$ & $A_{2u}+B_{1u}+E_{u}$ & $I$ \\
 \hline
lines &  &  &  &  \\
 \hline
$\Sigma $ & all & $A_{1g}+B_{1g}$ & $E_{u}$  & $I$\\
$\Delta $ & all & $A_{1g}+B_{2g}$ & $E_{u}$ & $I$ \\
$\Lambda $ & $A_{i}$, $B_{i}$ & $A_{1g}$ & $A_{2u}$ & $I$ \\
$\Lambda $ & $E$ & $A_{1g}+B_{1g}+B_{2g}+A_{1u}$ & $A_{2u}+B_{1u}+B_{2u}+A_{2g}$ & $I$\\
 \hline
points &  &  & &    \\
 \hline
Z & $ A_{i},$ $B_{i}$ & $A_{1g}$ & $-$ & $E$  \\
Z & $E$ & $A_{1g}+B_{1g}+B_{2g}$ & $A_{2g}$  & $E$\\
M & $t_{1}$, $t_{2}$ & $A_{1g}+B_{2g}+B_{2u}$ & $A_{1u}$  & $E$\\
M & $t_{3}$, $t_{4}$ & $A_{1g}+B_{2g}+A_{2u}$ & $B_{1u}$  & $E$\\
X & $t_{1}$, $t_{2}$ & $A_{1g}+A_{2u}+B_{1g}+B_{2u}+E_{u}$ & $E_{g}$ & $E$ \\
\hline Pair density waves $K=(b_{1}/2,b_{2}/2,0)$  &  &  &  &  \\
\hline
X & $t_{1}$, $t_{2}$ & $t_{2}+t_{4}$(M) & $t_{1}+t_{3}$(M) & $\sigma _{b}$ \\
    \end{tabular}
    \end{ruledtabular}
\end{table*}

In the present work we consider a single Cooper pair as a state of
two equivalent electrons. According to Anderson's ansatz
\cite{Anderson_PRB1984} the wavefunction of a Cooper pair is built
from one-electron states taking into account the Pauli exclusion
principle. Thus, for a generic point $k$ in the BZ, the spatial
parts of the wavefunctions for singlet and triplet pairs may be
written by the following two relations, respectively:
\begin{equation}
\psi _{k}^{s}
=\varphi_k(\xi_1)\varphi_{-k}(\xi_2)+\varphi_k(\xi_2)\varphi_{-k}(\xi_1)
 \label{a1}
 \end{equation}
 \begin{equation}
\psi^{t}_{k}
=\varphi_k(\xi_1)\varphi_{-k}(\xi_2)-\varphi_k(\xi_2)\varphi_{-k}(\xi_1)
\label{a2} \
\end{equation}

It is clear from the above formulas that the functions $\psi
_{k}^{s}$ and $\psi _{k}^{t}$ belong to IRs $A_{g}$ and $A_{u}$ of
group $C_{i}$, respectively.

The one-electron states in a crystal with a symmetry group $G$ are
defined by the wavevector $k$, its symmetry group $H$ (little group)
and the index $\kappa$ of small IR $D^\kappa$ of $H$ \cite{bc}. A
left coset decomposition of a space group $G$ with respect to $H$ is
written as:
\begin{equation}
G=\sum_\sigma s_\sigma H.
 \label{lcd}
\end{equation}

The action of the representatives of the left cosets $s_{\sigma }$ on
the wave vector $k$ results in a star $\left\{ k\right\} $ of the
wave vector. An IR of the space group $G$ is an induced
representation $D^{\kappa }\uparrow G$ defined by the following
equation \cite{bc}:
\begin{equation}
(D^\kappa \uparrow G)_{\sigma i \rho j} (g)=\left\{
\begin{array}{c}
D^\kappa _{ij}(s_\sigma^{-1}gs_\rho),\; \text{if
 }s_\sigma^{-1}gs_\rho \in H \\
0\text{, if }s_\sigma^{-1}gs_\rho \notin H
\end{array}%
\right. \text{,}
 \label{irsp}
\end{equation}
where we use Bradley's notation an up arrow for induction \cite{bc}.

The space group approach to the wavefunction of a Cooper pair
\cite{Yarzhemsky1992} is based on the Anderson representation
(\ref{a1}), (\ref{a2})  of a Cooper pair  and on standard
space-group representation theory for one-electron states
(\ref{irsp}). The  methods to join these two statements are
described elsewhere
\cite{Yarzhemsky1992,Teplyakov2025,Yarzhemsky2021} and the details
of the calculations are presented in  Appendix. In the following, we
briefly explain the physical background for the application of group
theoretical methods.

It is seen in Eqs. (\ref{a1}) and (\ref{a2})  that in agreement with
the Pauli exclusion principle the spatial part of a singlet
(triplet) pair is symmetrical (antisymmetrical) with respect to the
permutation of the electronic coordinates $\xi_1$ and $\xi_2$. Thus,
Anderson ansatz (\ref{a1}) may be generalized on the space groups by
constructing symmetrized and antisymmetrized Kronecker squares
$D^\kappa \uparrow G \otimes D^\kappa \uparrow G$  of space group
IRs (\ref{irsp}). Bearing in mind the induced structure of
space-group IRs one can use the Mackey-Bradley theorem on
symmetrized squares of induced representations
\cite{Yarzhemsky1992,Mackey1953,Bradley1970}. The complete Kronecker
square of dimension $\left\vert \hat{G}\right\vert ^{2}$ is
decomposed into subsets corresponding to the decomposition of the
space group into   double cosets relative to the wavevector group:
\begin{equation}
\mathbf{\ }G=\sum_{\delta }Hd_{\delta }H.
\end{equation}

Here and throughout $\hat{G}$ stands for the central extension of
the space group $G$, and the modulus sign denotes the number of
elements in a group. The wavevector $K_{\delta}$ of two-electron
state,   corresponding to the  double coset $d_{\delta}$ is defined
by the relation:
\begin{equation}
k+d_{\delta}k=K_{\delta}+b,
 \label{pdw}
\end{equation}
where $b$ is a reciprocal lattice vector.

The double coset defined by the inversion $I$ corresponds to a
Cooper pair:
\begin{equation}
k-k=0\text{ .} \label{cop}
\end{equation}

For a double coset defined by an identity element $E$ at the edges
of a BZ, the following relation can also be fulfilled:
\begin{equation}
k+k=b\text{ ,}
\end{equation}

In this case, two electrons with identical momenta are paired and
the equality of the total momentum to zero is achieved due to the
translational periodicity. If in Eq. (\ref{pdw}) $K_{\delta}=b_{i}/2$ the
lattice translation $T_{i}$ results in a phase factor $\pi$ for the
pair function and this case corresponds to PDW. Such case appears at
point $\mathrm{X}$, where $\mathrm{X+X'=M}$. In a generic point of a
BZ relation (\ref{cop}) holds and all even (odd) IRs are permitted
for singlet (triplet) pairs. but at symmetric points prohibitions on
some IRs arise. The results of decomposition of Kronecker  squares
for symmetrical points in a BZ for group $D_{4h}^{7}$ are presented
in Table~\ref{tab:table1}. It is seen in Table~\ref{tab:table1} that
at symmetry lines and planes some even (odd) IRs are forbidden. The
intersection of such line (plane) with a FS results in point (line)
node of Cooper pair wave function.

\begin{figure}
\includegraphics[width=0.85\columnwidth]{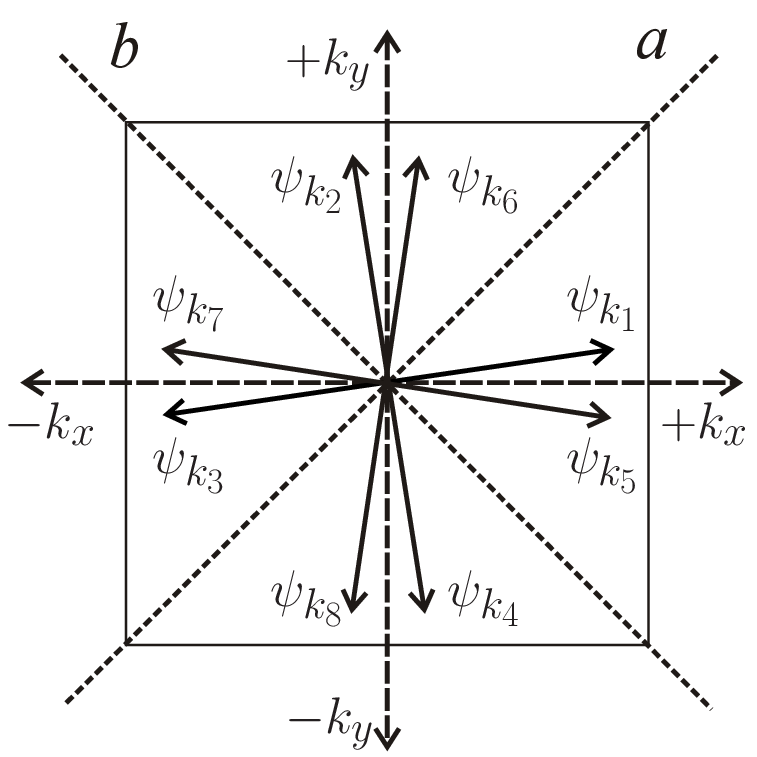}
\caption{Basis set of singlet and triplet pairs in $D_{4h}$ group in
a plane normal to $k_z$ direction. The $\psi_{k_i}$ denotes singlet
or triplet pair function defined by Eq. (\ref{a1}). Subscript $i$
corresponds to the action of the element of group $C_{4v}$ on the
initial vector $k_1$: $k_2=C_{4z}k_1$, $k_3=C_{2z}k_1$,
$k_4=C_{4z}^3$, $k_5=\sigma_y k_1$, $k_6=\sigma_a k_1$,
$k_7=\sigma_x k_1,k_8=\sigma_b k_1$.}
 \label{fig1}
\end{figure}

Anderson's Cooper pair wavefunctions (\ref{a1}) and (\ref{a2}) can
be generalized on a space-group symmetry by making use of the
standard projection operator technique for point groups, i.e. acting
on a single pair function by all elements of point group \cite{ham}.
For $k$ in a generic point of a BZ the pair's symmetry is $C_i$ and
it is sufficient to use just elements of group $\mathsf{C}_{4v}$. The basis
set of a Cooper pair for $k_1$ in the generic point of a BZ in
$D_{4h}$ symmetry is shown in Fig.~\ref{fig1}. The basis is chosen
in such a way that the action of elements of group $\mathsf{C}_{4v}$ results
in permutations of elements $\psi_{k_{i}}^{s(t)}$ without changing the
sign. Also, singlet functions are invariant with respect to the
inversion, while triplet functions change sign under the action of
inversion. These reducible basis sets can be easily decomposed into
irreducible sets belonging to IRs $\Gamma^q$ of point group
$\hat{G}$ making use of standard projection operator technique for
point groups \cite{ham}:
\begin{equation}
\Psi_i^{\Gamma^q}=\frac{|\Gamma^q|}{|\hat{G}^\prime|} \sum_{h_j \in
\bar{G}^\prime} \Gamma_{ii}^q (h_j) h_j\psi_{k_{1}}^{s(t)},
 \label{proj2}
\end{equation}
where $\hat{G}^\prime=\mathsf{C}_{4v}.$

\begin{figure}
\includegraphics[width=\columnwidth]{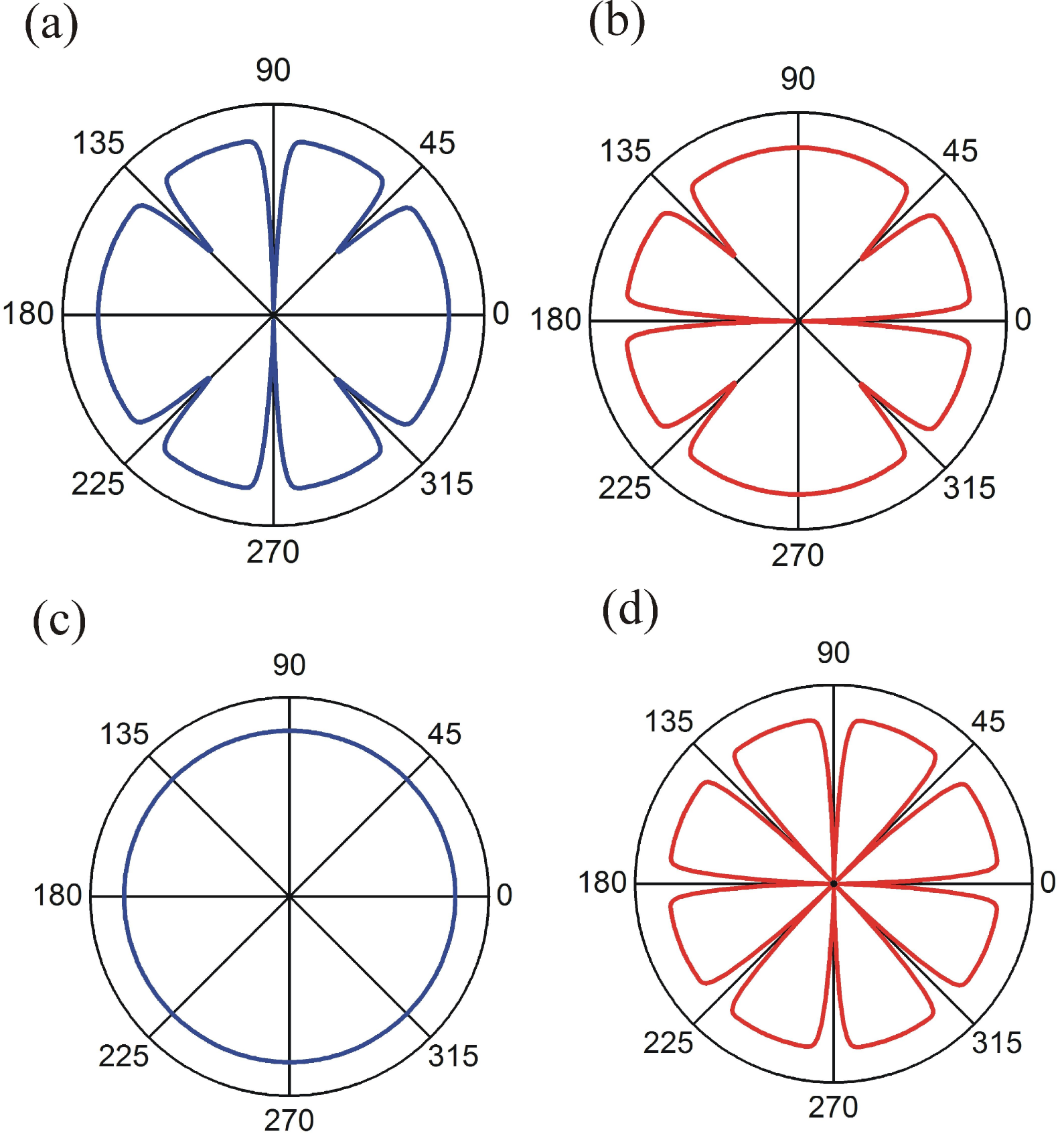}
\caption{\label{fig2} Nodal structure of the SOP in a plane normal
to $k_{z}$ direction. (a) $E_{g(u)}$ with $m=0$ and $m=1$, (b)
$E_{g(u)}^{A_{2g}}$ with $m=0$ and $m=1$, (c) $E_{1g(u)}^{\prime +}$
with $m=1 $, $E_{2g(u)}^{\prime +}$ with $m=-1$ and $B_{g(u)}^{+}$
with $m=2$, (d) the same as (c), but with $^{-}$ superscript. Note
that the nodal structures with $g$ and $u$ subscripts differ in the
basal plane (see Table \ref{tab:table1} and the text).}
\end{figure}

When a vector $k_1$ moves from the $k_x$ axis in Fig.~\ref{fig1} to
the $a$ axis, all other basis functions pass through the entire
circle. There is another free parameter - a phase winding
$\operatorname{exp}(im\theta)$     in the basis domain, where
$\theta$ is an angle between $k_1$ and $k_x$. Rotational elements
$\theta_j$ of point group that translate a vector $k_1$ into $k_2$,
$k_3$ and $k_4$ add additional phases:
\begin{equation}
\operatorname{exp}(i\bar {m} \theta_j)=
 \Gamma_{ii}^{q}(\theta_j).
 \label{expm}
\end{equation}

Thus each one-dimensional IR and each two-dimensional IR, reduced to
a diagonal form for rotations around a $k_z$ axis, can be associated
with a quasi-angular momentum $\bar{m}$. It is obvious that in order
for the phase winding to be the same for all values of $\theta$, the
quantities $m$ and $\bar{m}$ must be equated. In this way the SOP
(total  wavefunction of all pairs)  $\bar{\Psi} _{\Gamma^{q}
}^{s(t)}$  in the plane normal to $k_{z}$ direction is the
superposition of wavefunctions of all pairs belonging to IR
$\Gamma^{q}$ of point group $D_{4h}$, and may be written as:
\begin{eqnarray}
\bar{\Psi} _{\Gamma^{q} }^{s(t)}\left(\theta^{\prime}\right)=\sum_{\theta ,h_{j},i}\Gamma_{ii}^{q}(h_{j})h_{j}\psi ^{s(t)}_{k}\exp (im\theta ),\nonumber \\
\text{ }0\leq \theta <\pi /4,\text{ }  h_{j} \in \mathsf{C}_{4v}\text{,}
\label{sop}
\end{eqnarray}
where $\psi_{k} ^{s(t)}$ is the initial function of type (\ref{a1}) or
(\ref{a2}) with $k$ taken near $k_{x}$ axis ($\theta=0$). It is
clear from the above discussion that $\theta^{\prime}$ runs in the
interval from 0 to $2\pi$. In this representation a single pair is
characterized by quasi-angular momentum with respect to rotations of
point group, but continuous phase winding of SOP is developed in
pairs' condensate. Reflections interchange basis functions and add a
phase factor $-1$ in some cases, but do not affect a complex phase
$\operatorname{exp}(im\theta)$. It is clear (see Fig.~\ref{fig1})
that the directions of phase winding in adjacent sectors connected
by reflection are opposite to each other and the total phase winding
value equals to zero.  It can be easily shown that in order to
obtain a non-zero total phase winding, the mirror reflections of
$C_{4v}$ must be accompanied by complex conjugation (time reversal).
This corresponds to magnetic group symmetry. Since inversion is
already included in pair's symmetry, we come to the conclusion that
the unitary subgroup of $D_{4h}$ is $C_{4h}$ and magnetic group is
$4/mm^\prime m^\prime$.  Making use of Herring criterion (see e.g.
Ref.~\cite{bc}) we obtain that all ICRs (irreducible
corepresentations) of $4/mm^\prime m^\prime$ belong to type (a),
i.e. for unitary subgroup they are its IRs:
\begin{equation}
\Delta^\pm(R)=D(R),\;R\in C_{4h}. \label{icr1}
\end{equation}

They are extended into the non-unitary left coset defined by
$A=\vartheta\sigma_y$ as:
\begin{equation}
\Delta^\pm(B)=\pm D(BA^{-1}),\; B\in \vartheta C_{4h}.
 \label{icr0}
\end{equation}

The ICRs for group $4/mm^{\prime }m^{\prime }$ are given in
Appendix. The ICRs are denoted in the present paper by IRs of
unitary subgroup $C_{4h}$ with the sign in superscript,
corresponding to Eq. (\ref{icr0}). It is clear that the value of
quasi-angular momentum $\bar{m}$ of ICR is the same as that of IR of
the group $C_{4h}$  defined by Eq. (\ref{expm}). Eqs. (\ref{proj2})
and  (\ref{sop}) can be generalized for magnetic groups by adding
time-reversal $\vartheta$ to operators $h_{j}$ corresponding to
non-unitary elements. Since the operator $\vartheta$ is antilinear
\cite{bc}, reflections in vertical planes are accompanied with
complex conjugation and the direction of the phase winding is the
same in all sectors. In this case total phase winding is $2\pi m$.
Since ICRs with $+$ and $-$ sign are possible, a phase difference on
vertical plane may be zero or $\pi $. This means that, states with
and without nodes are possible. In the present approach the SOP is
constructed as a superposition of wavefunctions of all pairs, but
its nodal structure is defined by the IR of a single pair. For
example, when $k_{1}$ in Fig. \ref{fig1} approaches $k_{x}$ axis the
pair function $\psi_{k_{1}}$ can cancel or merge with the pair function
$\psi_{k_{5}}$. Which of the two cases is realized is determined by the
sign of matrix element of IR $\Gamma ^{q}$ of point group in Eq.
(\ref{proj2}) for the reflection element. The plus and minus signs
correspond to nodeless and nodal cases, respectively.

\subsubsection{Spin part of pair function}

In the axial symmetry triplet spin part $(\hat{x},\hat{y},\hat{z})$
splits into the OSP (opposite spin pairing) part $S_{0}^{1}=\hat{z}$
and the ESP (equal spin pairing) part $(\hat{x},\hat{y})$. For our
purposes, it is more convenient to use complex linear combinations
that have certain values of projection of the angular momentum onto
the $z$ axis $S_{\pm 1}^{1}=\hat{x}\pm i\hat{y}$. In the symmetry
$D_{4h}$ two components $\hat{x}\pm i\hat{y}$  belong to the complex
IR $E_{g}$ and the $\hat{z}-$component belongs to IR $A_{2g}$. The
characters of $A_{2g}$ coincide with that of ICR $A_{g}^{-}$(see the
appendix). This function is nodeless in the basal plane and has
nodes in all vertical planes.  Consider transformation properties of
the $\hat{x}+i\hat{y}$  component in the symmetry planes under the
action of the elements of the point    group:
\begin{equation}
\sigma _{x}\left( \hat{x}+i\hat{y}\right) =\hat{x}-i\hat{y}.
\end{equation}

Hence, it follows that reflections change the spin projection. On
the other hand, the reflection $\sigma_{x}$ with time reversal does
not change the spin projection:
\begin{equation}
\vartheta \sigma _{x}\left( \hat{x}+i\hat{y}\right) =\theta \left( \hat{x}-i%
\hat{y}\right) =\hat{x}+i\hat{y}.
\end{equation}

Reflection $\sigma_{y}$ with time reversal does not change the spin
projection, but adds a phase factor $-1$:
\begin{equation}
\vartheta \sigma _{y}\left( \hat{x}+i\hat{y}\right) =\theta \left( -\hat{x}+i%
\hat{y}\right) =-\hat{x}-i\hat{y}.
\end{equation}

Hence, it follows that the spin function $\uparrow \uparrow =\left(
\hat{x}+i\hat{y}\right) $ belongs to the ICR $E_{1g}^{\prime -}$ of
the magnetic group $4/mm^{\prime }m^{\prime }$. It can be shown by a
similar way that the spin function $\downarrow \downarrow =\left(
\hat{x}-i\hat{y}\right)$ belongs to ICR $E_{2g}^{\prime -}$. The
action of the reflection in the basal plane changes the sign of the
ESP spin function:
\begin{equation}
IC_{2z}\left( \hat{x}+i\hat{y}\right) =-\hat{x}-i\hat{y},
\end{equation}
which means that it is nodal in the basal plane.

\subsubsection{Numerical calculation of SOP}
In our approach SOP is a superposition of wavefunctions of all
pairs. Since we are interested in the nodal structure in vertical
planes, it is sufficient to consider the intersection of the FS with
a plane perpendicular to the $k_{z}-$ axis, which is approximated by
a circle. In the case of phase winding with $m=1$ in $D_{4h}$
symmetry, the phase difference between two sides of the vertical
reflection plane may be complex. In this case, the structure of
squared SOP may be obtained by computer modeling as follows. At 360
points on a circle, representing the intersection of FS with a plane
normal to the $k_{z}$ direction, the real and imaginary parts were
modeled by Gaussians with a small half-width and complex
coefficients corresponding to the theoretical phase. At each point,
the sums of contributions of all pairs to the real and imaginary
parts and the square of the modulus of this complex number were
calculated. The square of the superposition of all pair's
wavefunctions has the same physical meaning as the square of pair's
function (which is identified with SOP) in the Ginzburg-Landau
functional. The magnitude of the projection of the angular momentum
$m$ for the phase winding was chosen according to the Eq.
(\ref{expm}), i.e. $m=1$ for IRs and ICRs of $E-$ type and $m=2$ for
IRs and ICRs of $B-$ type. The results are presented in Fig.
\ref{fig2}.

\begin{figure*}
  \includegraphics[width=0.9\textwidth]{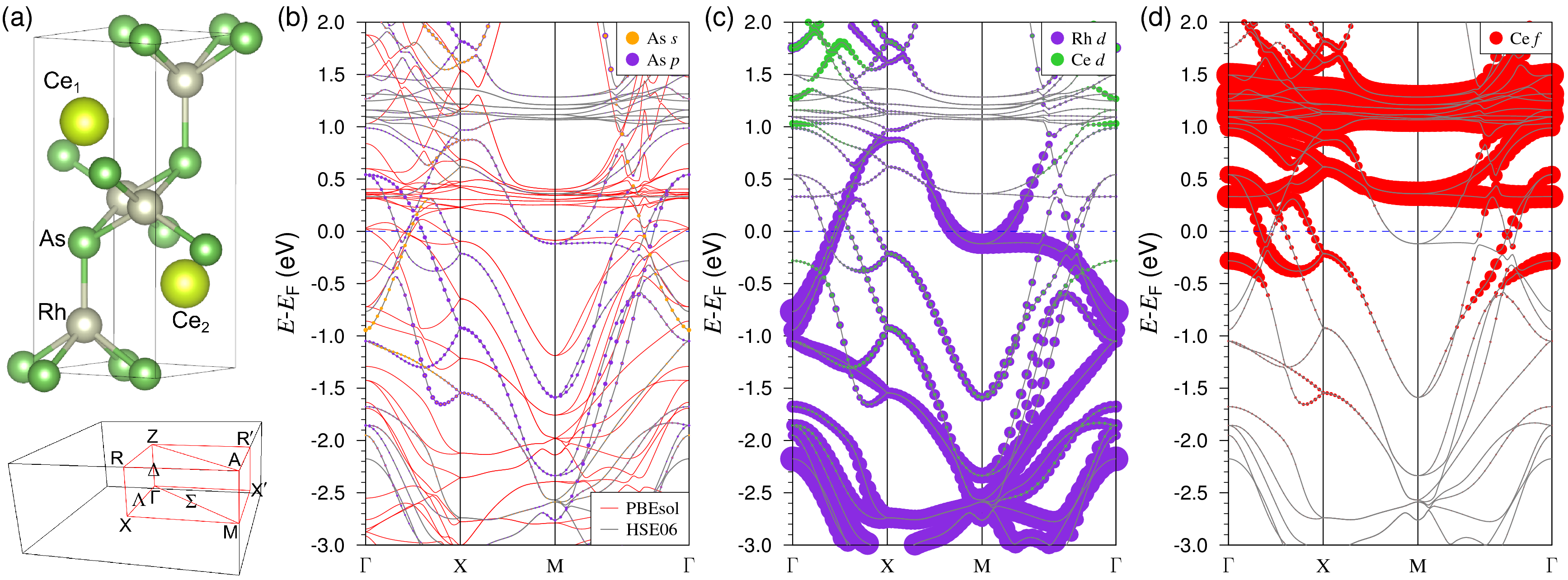}
  \caption{(a) Tetragonal CaBe$_2$Ge$_2$-type atomic structure of CeRh$_2$As$_2$ (top) and corresponding bulk Brillouin zone (bottom) with indication of the high-symmetry points and high-symmetry directions. The bulk electronic band structure of CeRh$_2$As$_2$ calculated with the GGA-PBEsol and the HSE06 hybrid functional with weights of the As-$s,p$ (b), Rh(Ce)-$d$ (c), and Ce-$f$ (d) orbitals.}
  \label{fig_bulk_HSE}
\end{figure*}

\begin{figure*}
  \includegraphics[width=0.9\textwidth]{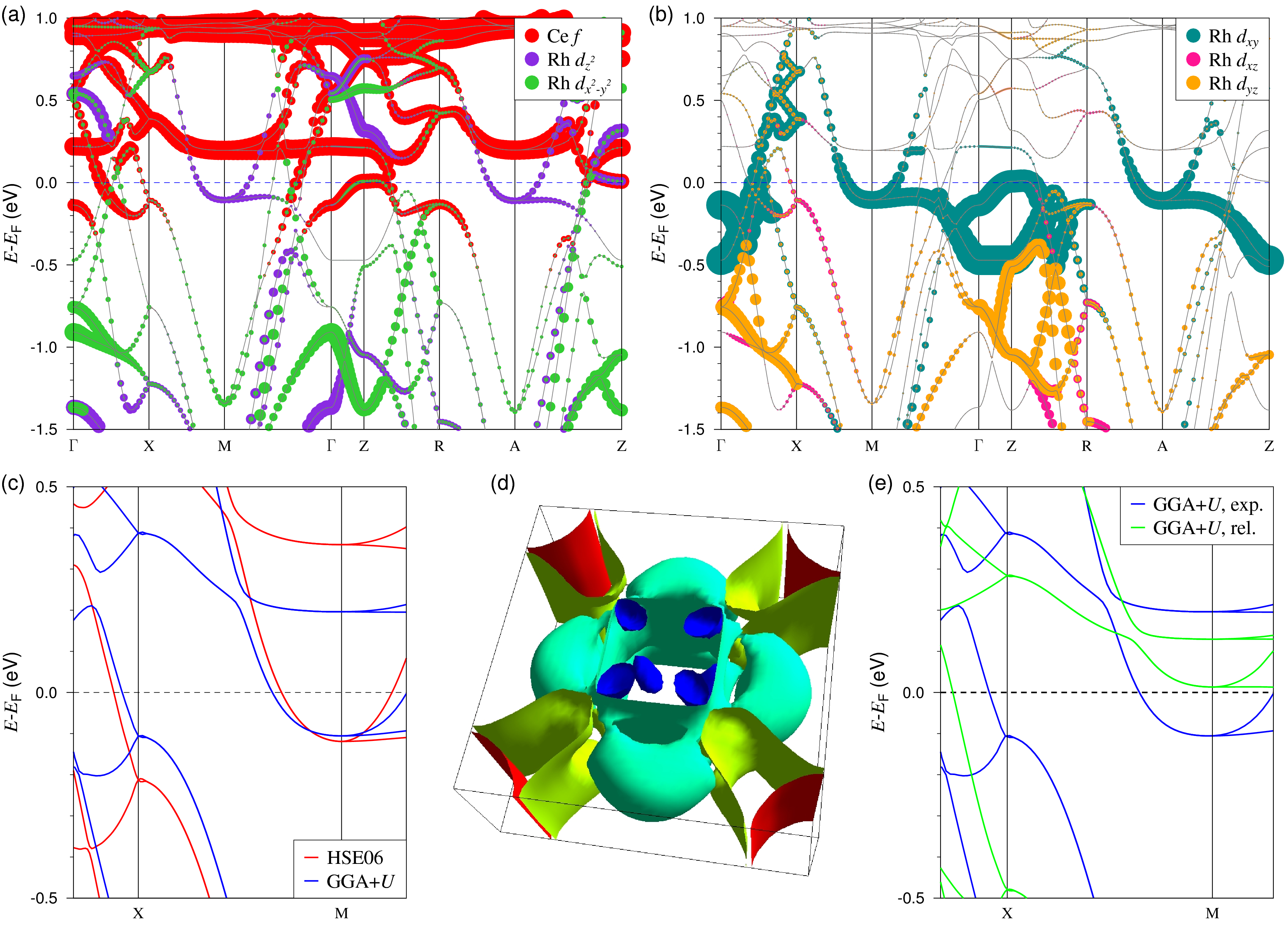}
  \caption{Bulk electronic band structure calculated within GGA+$U$: (a) Ce-$f$ and Rh-$e_g$ ; (b) Rh-$t_{2g}$ orbitals. (c) Magnified view of the GGA+$U$ band structure in the vicinity of X and M points in comparison with HSE06 result. (d) Calculated Fermi surface. (e) Comparison of low-energy spectra calculated with experimental and relaxed atomic structure parameters.}
  \label{fig_bulk_GGA+U}
\end{figure*}

\subsection{DFT calculations}

Electronic structure calculations were carried out within the
density functional theory using the projector augmented wave (PAW)
method \cite{Blochl.prb1994} as implemented in the VASP code
\cite{vasp1,vasp2}. To obtain accurate bulk band structure, the
HSE06 screened hybrid functional \cite{HSE06} was adopted. Since
hybrid functional approach is resource consuming for Fermi surface
calculations the correlation effects were included within the
GGA+$U$ method~\cite{Anisimov1991}. The values of the screened
Coulomb $U$ and Hund $J$ interaction parameters for Ce $4f$
electrons were obtained variationally to fit the $4f$ bands position
calculated with HSE06. The Fermi surface was determined on a dense
$24 \times 24 \times 10$ $k$-point mesh and visualized by using
\texttt{FermiSurfer} \cite{Kawamura2019}.

\section{Results}

\subsubsection{Electronic structure}

Previous DFT calculations \cite{nogaki,Ishizuka_PRB2024}  resulted
in some intersections of FS with faces of a BZ. The pairing of
electrons  with $k$ vectors on the  faces of a BZ can violate the
direct connections between multiplicity and parity of a pair that
follow from the Anderson ansatz (\ref{a1}), (\ref{a2}) in generic
$k$-points \cite{Yarzhemsky1992}. In order to estimate the
contribution of Ce $4f$ electrons in symmetry points of a BZ, new
more sophisticated DFT calculations were carried out in the present
work.

We begin with a standard DFT with the PBEsol exchange-correlation
functional \cite{PBEsol}. As a starting point, we use experimental
crystal structure parameters (both lattice constants and atomic
positions, Fig.~\ref{fig_bulk_HSE}(a) adopted from
Ref.~\cite{khim1}. The resulting bulk band structure
(Fig.~\ref{fig_bulk_HSE}(b), red lines) partially reproduces earlier
calculations \cite{nogaki,Ptok_PRB2021,Wu_CPL2024,Chen_PRX2024} were
performed without and with the Ce-$4f$ correlations. The latter was
incorporated within the GGA+$U$ with a relatively small Hubbard
parameter ($U=2$ eV). Across all calculations, including ours, the
position of the Ce-$4f$ band was found to be slightly above the
$E_\mathrm{F}$. The difference from earlier calculations is the
appearance of an electron band intersecting the $E_\mathrm{F}$ at
the M point, which was unoccupied in previous works. In the
following, we will show that this difference is related to lattice
relaxation, which we do not include yet here.

The spectrum calculated with the screened hybrid functional HSE06,
which more accurately depicts the correlation effects, shows that
the position of the dispersionless unoccupied $f$-band is much
higher, above $+1$ eV (Fig.~\ref{fig_bulk_HSE}(b), gray lines). The
change in the position of the $f$ band does not affect the M point
band while it essentially modifies the dispersions of the states
along the $\Gamma - {\mathrm X}$ and $\Gamma - {\mathrm M}$
directions near the Fermi level. This distinction in the effect of
the $f$ band position on the states at the Fermi level is rooted in
their distinct orbital compositions. The electronic state at the
$\mathrm M$ point is completely composed of the $d$ electrons of
rhodium (Fig.~\ref{fig_bulk_HSE}(c)) whereas the $\Gamma$-centered
hole-like states involve a significant $f$ contribution
(Fig.~\ref{fig_bulk_HSE}(d)). Note that the contribution of arsenic
orbitals to the states near $E_{\mathrm F}$ is negligibly small
(Fig.~\ref{fig_bulk_HSE}(b)).  This analysis underscores the
importance of employing advanced functionals  such as HSE06 for
precise predictions of electronic structure, particularly when
subtle orbital effects influence material properties.

Next we turn to a simplified approach, namely GGA+$U$, adjusting the
Hubbard $U$, and Hund $J$ interaction parameters for cerium $f$
orbitals to match the position of the $f$ band obtained with the
HSE06 functional. We found that at $U$ up to 5 eV the position of
the $f$ band remains almost unchanged, approximately at $+0.5$ eV,
similar to the GGA calculation. Only with Hubbard $U=10$~eV the
position of the $f$ band reaches $+1$ eV
(Fig.~\ref{fig_bulk_GGA+U}(a)).  This also requires a rather large
parameter $J$, 2 eV, because there are convergence problems with
smaller values. Overall, the obtained GGA+$U$ band spectrum agrees
well with the HSE06 result. Figs.~\ref{fig_bulk_GGA+U} (a,b) show
the detailed symmetry of Rh $d$ states along high-symmetry
directions in the bulk Brillouin zone. In particular, the M band,
which also persists at $k_z=0.5$ (the A point), is primarily
composed of Rh $d_{xy}$ orbitals with a minor $d_{z^2}$
contribution. It  should be noted that the M band is well resolved
in ARPES \cite{Chen_PRX2024}. Its energy and dispersion are
described within GGA+$U$ with an accuracy of HSE06
(Fig.~\ref{fig_bulk_GGA+U}(c)) and since it does not exhibit a $k_z$
dependence it leads to the formation of cylindrical sheets of the
Fermi surface, centered at the corners of the Brillouin zone
(Fig.~\ref{fig_bulk_GGA+U}(d)).

The next state we will focus on is the band at the X point, which is
also observed in ARPES and its analysis has been the subject of
previous papers \cite{Wu_CPL2024,Chen_PRX2024}. High-resolution
ARPES measurements revealed a band with an electron-like dispersion
along $\Gamma-\mathrm{X}-\Gamma$ and a hole-like dispersion along
the $\mathrm{M}-\mathrm{X}-\mathrm{M}$ direction, indicating the
existence of a van Hove singularity (VHS) at the X point. The
binding energy of the VHS was found to be approximately $75\pm 60$
meV \cite{Chen_PRX2024}. Notably, our GGA+$U$ calculations place the
X-band $\approx 100$ meV below $E_\mathrm{F}$, agreeing well with
experiment, unlike HSE06 calculations, in which the VHS lies a
hundred meV lower (Fig.~\ref{fig_bulk_GGA+U}(c)).

Finally, we examined the effect of structural relaxation on the
low-energy bulk spectrum. We found that the relaxation only subtly
altered the lattice parameters of CeRh$_2$As$_2$, contracting them
by roughly $\approx 1$~\%, specifically $a=b$ by 0.91~\% and $c$ by
1.04~\%, compared to their experimental values \cite{khim1}.
However, it significantly modifies interlayer distances within the
cell: the relaxed Rh--As interlayer distances within Rh-As-Rh and
As-Rh-As trilayers shortened by 12.7~\% and expanded by 14.8~\%,
respectively, relative to the interlayer distances, derived from the
accurate experimentally measured atomic coordinates \cite{khim1}.
Along with this, Ce-Rh and Ce-As interlayer distances also undergo
significant relaxation displacements, by $+6.9$~\% and $-10.5$~\%,
respectively. These substantial atomic structural changes, despite
negligible changes in the lattice parameters, strongly modify the
band structure (Fig.~\ref{fig_bulk_GGA+U}(e)). The occupied band at
the M point becomes depleted, consistent with prior calculations
\cite{nogaki,Ptok_PRB2021,Wu_CPL2024,Chen_PRX2024}, where the
lattice relaxation was also taken into account. At the same time,
the VHS state at the X point after relaxation lies much lower,
approximately at 0.5 eV below $E_\mathrm{F}$, contradicting ARPES
results ($75\pm 60$ meV \cite{Chen_PRX2024}). Therefore, DFT
relaxation appears to induce unrealistically large internal atomic
distortions and the electronic band spectrum of such a relaxed
structure contradicts experimental observations.

\subsubsection{Spatial part of pair function   inside the BZ and on its faces}

Possible IRs for singlet and triplet pair functions on symmetry
points, lines, and planes in a BZ obtained by the Mackey-Bradley
theorem are presented in Table \ref{tab:table1}. The absence of any
IR on a symmetry line or plane means group-theoretical point or line
node of the pair function belonging to this IR. When comparing the
data of Table \ref{tab:table1} with the characters of IRs of
$D_{4h}$ presented in the Appendix we see that the negative
characters of the one-dimensional IRs in all planes and of
two-dimensional IR in basal plane correspond to nodal planes.
Characters of $E_{g(u)}$ in vertical planes equal to zero and nodes
depend on the unitary transformation. These nodes and dips will be
considered later by numerical calculations. Due to the
non-symmorphic structure of the space group, nontrivial pairing
symmetry appears on the faces of a BZ. On the faces of a BZ normal
to $k_{y}$  (plane X$^\prime$MAR$^{\prime}$, see Fig.~\ref{fig_bulk_HSE} (a, bottom))
singlet $A_{1g}$ pairs are forbidden. At points M and X, singlet odd
pairs are possible, and at points X and Z, triplet pairs are even.
Also, at point X, coupling of two electrons with $k=b_{1}/2$ and
$k=b_{2}/2$ results in PDW pairs with $K=(b_{1}/2,b_{2}/2,0)$. In
this case, a gap function changes the sign under the action of
single lattice translations $t_{1}$ or $t_{2}$.

Now let us move on to the consideration of two-dimensional IRs at
generic points of a BZ, where all even (odd) IRs are possible for
singlet (triplet) pairs. It is immediately verified that a
multiplication of $E_{g(u)}$ by $A_{2g}$ results in also  IR
$E_{g(u)}$, which we denote by the superscript $A_{2g}$. Since the
characters of $A_{2g}$ equal to $-1$ for reflections in vertical
planes, this transformation reverses the nodal structure in vertical
planes. Inside the sectors, the interference is almost constructive
and the squared SOP is constant. Interference of pair's functions
located on both sides of the symmetry plane can lead to the
appearance of nodes and dips  of the squared SOP.  The matrices of
$\tilde{E}_{g}$ and $\tilde{E}_{u}$ in diagonal complex form for
rotations $\mathsf{C}_{4}$ are presented in the Appendix. Since the matrices
of $\tilde{E}_{g}$ and $\tilde{E}_{u}$ are chosen in such a way that
they coincide on the subgroup $\mathsf{C}_{4v}$, the nodal structures of SOP
in vertical planes in Figs. \ref{fig2} (a) and (b) are the same for
$\tilde{E}_{g}$ and $\tilde{E}_{u} $. The structure in Fig.
\ref{fig2} (b) corresponds to the multiplication of the IR
$\tilde{E}_{g}$ or $\tilde{E}_{u}$ by $A_{2g}$. Computer
calculations showed that the structures  without phase winding
and with  phase winding with $m=\pm 1$ are the same. Both structures
in Figs. \ref{fig2} (a) and (b) are characterized by dips in both
diagonal planes. Structure in Fig. \ref{fig2} (a) is nodeless, but
the structure corresponding to IR $\tilde{E}_{g}$ multiplied by
$A_{2g}$ has nodes in planes normal to the axes $k_{x}$ and $k_{y}$.
Thus, the nodes of $\tilde{E}_{g}$ and $\tilde{E}_{u}$ in vertical
planes are topological, i.e., they can be removed by a unitary
transformation that does not change the characters. Nodes in the
basal plane do not depend on unitary transformations and they are
the same as those obtained using the Mackey-Bradley theorem (see Table
\ref{tab:table1}), namely $\tilde{E}_{g}$ is nodal and
$\tilde{E}_{u}$ is nodeless in the basal plane. Calculations showed
that nodes of one-dimensional IRs $B_{1(2)g(u)}$ in the chiral case
are the same as those that follow from group theory (see Table
\ref{tab:table1}) and they are not shown in Fig. \ref{fig2}. Since
in the chiral case, the phase winding directions in the sectors
connected with reflections are opposite, total phase winding is
zero. Basis functions of ICRs can be easily constructed using Eq.
(\ref{proj2}) by replacing $h_{j}$ with $h_{j}\vartheta $ for
anti-unitary elements. Thus, we can write the expression for the ICR
originating from the IRs $E_{1g(u)}^{ \prime \pm}$ of $C_{4h}$:
\begin{eqnarray}
\Psi \left( E_{1g(u)}^{\prime \pm }\right) =(\psi _{k_{1}}^{s(t)}+i\psi
_{k_{2}}^{s(t)}-\psi _{k_{3}}^{s(t)}-i\psi
_{k_{4}}^{s(t)})e^{i\theta }  \label{icre} \notag \\
\pm ( \psi _{k_{5}}^{s(t)}-i\psi _{k_{6}}^{s(t)}--\psi _{k_{7}}^{s(t)}+i\psi
_{k_{8}}^{s(t)})e^{-i\theta}.
\end{eqnarray}

Note that since reflections change  the direction of the contour
traversal, the phase winding direction  is the same in all sectors.
For example, the element $\vartheta\sigma_{y}$ transforms the two
points $\theta=0$ and $\theta=\pi/4$ with phases $\phi=0$ and
$\phi=\pi /4$ into the points $\theta=0$ and $\theta=-\pi/4$ with
phases $\phi=0$ and $\phi=-\pi/4$ and the phase winding direction
remains unchanged.

In magnetic group symmetry, the nodal structures of SOP in vertical
planes are the same for all ICRs with a plus sign and for all ICRs
with a minus sign (see Figs.  \ref{fig2} (c) and (d), respectively).
Note that in numerical calculations  $m=1$  for $E_{g(u)}^{\prime
\pm}$  and  $m=2$ for $B_{g(u)}^{\pm}$ were used.  However, in the
magnetic symmetry $4/mm^{\prime }m^{\prime }$ the total phase
winding is $2\pi m$ in both nodeless and nodal cases.
 In
the case of the magnetic group $4/mm^{\prime }m^{\prime }$ with
phase winding $m=1$ for $E_{1g(u)}^{\prime +}$ and $m=2$ for \
$B_{g(u)}^{+}$, the resultant SOPs have no nodes in vertical planes,
but SOPs $E_{1g(u)}^{\prime -}$ and $B_{g(u)}^{-}$ have nodes in all
vertical planes, as shown in Figs. \ref{fig2} (c) and (d),
respectively.

Thus, we have obtained that with axial symmetry, the nodes on
vertical symmetry planes can be removed in magnetic symmetry with
phase winding, if the  phase winding parameter $m$ corresponds to
the character of the IR (see Eq. (\ref{expm})). At the same time,
the nodes on the basal plane are uniquely related to the projection
of the angular momentum and are robust. Indeed, for $C_{2z}$
rotation characters equal to $-1$ and $+1$ for $m=1$ and $m=2$,
respectively. Since reflection in the basal plane equals to
$IC_{2z}$, we obtain for $m=1$ that $\chi_{g} (\sigma _{z})=-1$ and
$\chi_{u} (\sigma _{z})=+1$ for even and odd pairs, respectively.
Similarly, at $m=2$, we obtain that $\chi_{g} (\sigma _{z})=1$ and
$\chi_{u} (\sigma _{z})=-1$. Thus, robust nodes in the basal plane
appear at $m=1$ and for even pairs and at $m=2$ for odd pairs. And
vice versa: even pairs with $m=2$ and odd pairs with $m=1$ are
nodeless in basal plane.

\subsubsection{Total pair function inside the BZ}

It follows from the above discussion that in a singlet case there
are two nodeless states, corresponding to experimental data for the
SC1 phase \cite{juraszek2025}, namely ICR $B_{g}^{+}$ with phase
winding $m=2$ in the magnetic group $4/mm'm'$ symmetry  and  IR
$A_{1g}$ without phase winding in $D_{4h}$ symmetry.

In a triplet case nodeless spatial part  in $4/mm^{\prime }m^{\prime
}$ symmetry belongs to ICR  $E_{1u}^{\prime +}$ with phase winding
$m=1$. The total pair function is a direct product of its spatial
and spin parts. In axial symmetry, the triplet spin function splits
into  ESP and OSP  parts belonging to $E_{1g}^{\prime -}\left(
E_{2g}^{\prime +\ \ }\right) $ and $A_{g}^{-} $ respectively.
Multiplyng the nodeless spatial part of a triplet pair
$E_{1u}^{\prime +}$ by $A_{g}^{-}$ results in nodal $E_{1u}^{\prime
-}$. Although   when multiplying the $E_{1u}^{\prime -}$ spatial
part by the spin part $A_{g}^{-}$ we obtain the nodeless total pair
function $E_{1u}^{\prime +}$. In $D_{4h}$ symmetry, multiplication
of the ESP spin part $\tilde{E}_{g}$ (nodal in the basal plane) by the
triplet spatial part $\tilde{E}_{u}$ (nodeless in the basal plane)
results in IRs $A_{1u}+A_{2u}+B_{1u}+B_{2u}$, which are nodal in the
basal plane. This coupling of spatial and spin parts of a Cooper
pair is similar to angular momentum coupling in atomic physics.
Indeed two lines of two-dimensional IRs correspond to angular
momentum projections $+1$ and $-1$ and in the direct product there
are terms $m=2$, $m=-2$, and two terms with $m=0$, which correspond
to $B_{1u}$, $B_{2u}$, $A_{1u}$ and $A_{2u}$.

Similar angular momentum coupling takes place in $4/mm^{\prime
}m^{\prime}$ symmetry. When multiplying the ESP function
$E_{1g}^{\prime-}$ by the spatial part $E_{1u}^{\prime -}$ or
$E_{2u}^{\prime -}$,  we obtain $B_{u}^{+}$ or $A_{u}^{+}$,
respectively. There is some similarity and some difference between
our results and the results of  the order parameter modeling with
spherical functions \cite{nogaki2}. Our functions $E_{1u}^{\prime
+}$ or  $E_{2u}^{\prime +}$ are similar to functions $E_{u}^{\prime
}$ and $E_{u}^{\prime \prime }$ \cite{nogaki2}; however, our helical
functions $B_{u}^{+}$ may have angular momentum projection $m=2$ and
phase winding $4\pi m$, but the angular momentum of $B_{u}$ is not
expressed explicitly \cite{nogaki2}.  The model approach
\cite{nogaki2} resulted in a fully gapped $A_{u}$ in $C_{4h}$
symmetry. In the present space-group approach, the IR $A_{1u}$ of
the $D_{4h}$ group has nodes in all planes (see Table
\ref{tab:table1}). The nodes in vertical planes of $A_{1u}$ may be
removed by by consideration of $A_{u}^{+}$ with the phase winding
$m=4$, but, as was shown above, nodes on the basal plane of an odd
function with $m=4$ are robust. Recent results of local
magnetization measurements indicated fully-gaped superconductivity
\cite{juraszek2025} in the SC1 phase and staggered magnetization in
the basal plane. This may correspond to singlet $A_{1g}$ pairs.  We
obtained that there are two additional options for fully gapped
triplet functions both of symmetry $E_{1u}^{\prime +}$. This ESP
state is consistent with a $d-$vector in the $ab-$plane
\cite{Landaeta_PRX2022}. The OSP case corresponds to the triplet
spatial part $E_{1u}^{\prime -}$with $m=1$, and in the ESP case to
the triplet spatial part  $B_{u}^{-}$ with $m=2$. Note that despite
the same full symmetry $E_{1u}^{\prime +}$ and total angular
momentum, the structure of these pairs is different. In the OSP
pair, the electron spin projections are opposite and the angular
momentum is determined by the spatial part. In this case, the
magnetic field leads to the destruction of pairs. In the ESP case,
the angular momentum can be determined only by the spin or  by the
sum of the angular momentum of the spin and spatial part. In this
case, the magnetic field does not destroy the pair, but rather
maintains the electron spins in one direction.  Also, a singlet
$A_{1g}$  is nodeless and can't be ruled out for a low-field state.

\section{Discussion}

The above considered group theoretical results may be used to build
models of parity changing transition without changing of pairing
potential. It is very remarkable that the crystal structures of the
heavy fermion superconductors containing $d$ and $f$ electrons,
e.g., UPt$_{3}$, UBe$_{13}$, LiPt$_{3}$B, and also of recently
discovered UTe$_{2}$ and CeRh$_{2}$As$_{2}$, are non-symmorphic
\cite{Micklitz_PRL2017}. Below, we will shed light on the
relationship between superconductivity and non-symmorphic structure
within the framework of the nearest-neighbor exchange interaction
model.

There are two nonequivalent Ce atoms per unit cell, denoted as
Ce$_1$ and Ce$_2$ (see Fig.~\ref{fig_bulk_HSE}(a)), and there is no
inversion symmetry at either Ce atom. However, the two nonequivalent
Ce atoms are inversion symmetry partners of each other, so there is
global inversion symmetry.  For $k$ being a generic point in the BZ,
the IR of the space group $D_{4h}^{7}$ is defined by a
16-dimensional wavevector star (see Eq. (\ref{irsp})). Half of the
wavevector star is shown in Fig.~\ref{fig1}. This set of prongs,
which we call upper set, is invariant with respect to the point
group $\mathsf{C}_{4v}$, and another half of the prongs (lower set, which is
not shown in Fig.~\ref{fig1}) is obtained by the action of the
element $\left\{ I|\tau\right\}$. Improper translation $\tau$ is not
essential for the construction of IRs of the space group inside the
BZ, but it is important for construction of the basis set.

The basis set belonging to an IR of a space group may be constructed
from atomic states, making use of projection operators (see Eq.
(\ref{proj2})), where the sum runs over all point group elements and
all lattice translations $T$. The action of the elements of $\mathsf{C}_{4v}$
on the initial $k$ vector  results in the prongs of the upper set
shown in Fig.~\ref{fig1}. Since the local subgroup of  Ce$_1$ $4f$
functions is $\mathsf{C}_{4v}$, they are projected on the $k$ vectors of the
upper set. Since the translation subgroup $T$ belongs to the
wave vector group, the basis functions of all sublattice
Ce$_1$$\times T$ are projected on the prongs of the upper face. The
elements of the left coset $\left\{ I|\tau\right\} \mathsf{C}_{4v}$
interchange sublattices and also the prongs connected by space
inversion.  Due to exchange interaction between nearest neighbors
the lowest energy will correspond to the triplet term of electrons
on two Ce atoms with antisymmetrized spatial part. According to the
group theory, the scalar interaction is possible between the states
belonging to the same row of the same IR only.  As we demonstrated
above the wavefunctions of two Ce atoms connected by $\left\{ I|\tau
\right\}$ belong to opposite $k$ vectors, i.e. different rows of IR,
and hence, the  exchange interaction between them is symmetry
forbidden. In a Cooper pair the opposite $k$ vectors cancel each
other ($K_{\delta}=0$) and hence the exchange interaction between
nearest neighbors  is not restricted by the symmetry. On the other
hand, in symmorphic groups equivalent atoms are connected by
translations belonging to $T$. Since $T$ belongs to the wavevector
group in all cases, the atomic functions of nearest neighbors belong
to the same $k$ vector, i.e. to the same row of IR and hence, the
exchange interaction between them is possible without Cooper
pairing.  Thus, the exchange interaction of electrons on two nearest
neighbors atoms connected by $\left\{ I|\tau \right\} $ can cause
Cooper pairing in non-symmorphic space groups only.

To acquire the parameter of the model,  we evaluated the exchange
integral for $4f$ electrons on Ce$_1$ and Ce$_2$ atoms located at an
interatomic distance of 5.7 \r{A}, using Ce Hartree-Fock atomic
functions, and determined that the exchange integral
$G(4f_{1},4f_{2})$ equals $2.7 \times 10^{-5}$ eV.  This value
corresponds to the order of magnitude of  $T_{c}$ of
CeRh$_{2}$As$_{2}$ \cite{khim1}. Hence, it follows that our model of
Cooper pairing due to exchange interaction between two Ce atoms
connected by the element  $\left\{I|\tau\right\}$ is energetically
possible.

Earlier the spin-singlet AFM superconductivity was proposed on the
base of the NMR measurements~\cite{Ogata_PRL2023}. However, it
should be noted that, strictly speaking, NMR may distinguish between
even and odd spatial functions and this result  may correspond to
OSP triplet pairs at the X point, which are even (see
Table~\ref{tab:table1}). According to our DFT calculations
(Fig.~\ref{fig_bulk_HSE}(d)), the bands with a large contribution of
Ce-$4f$ electrons, intersects the Fermi level in the
$\Gamma-\mathrm{X}$ direction near the X point. Thus, the SC1 state
can be explained in terms of even triplet OSP pairing at X.
Antiferromagnetism is no longer observed upon the parity transition
to the high-field SC2 state \cite{Ogata_PRL2023}. Within our model
this behavior can be connected to odd ESP triplet pairs in the
$\Gamma-\mathrm{X}$ direction.

Thus, the proposed scenario, based on space group symmetry, and DFT
calculations can explain the symmetry changing transition in
CeRh$_{2}$As$_{2}$ within a single pairing potential, namely
nearest-neighbor exchange interaction of Ce-$4f$ electrons. Our
findings also allow us to propose a parity-change scenario for
singlet pairs. Indeed, our DFT and ARPES \cite{Wu_CPL2024} results
have shown the presence of the Fermi surface pockets centered at the
M point (Fig.~\ref{fig_bulk_GGA+U}(d)), which are populated by
Rh-$4d$ orbitals (Fig.~\ref{fig_bulk_GGA+U}(a,b)). The VHS state at
X below $E_\mathrm{F}$ is contributed both Ce-$4f$ and Rh-$4d$
orbitals. Given the non-symmorphic structure of the space group,
singlet pairs at the X  and M points can be either even or odd (see
Table~\ref{tab:table1}). Consequently, it follows that the even-odd
transition can be described as, for example, an $A_{1g}\rightarrow
B_{2u}$ transition of singlet pairs at symmetry points on the faces
of a BZ.

Very recently, it has been discovered that the superconductor
LaRh$_2$As$_2$, which shares the same crystal structure as
CeRh$_2$As$_2$ and has a similar critical temperature ($\approx 0.3$
K), exhibits conventional full-gap $s$-wave type-II
superconductivity. This is characterized by a small upper critical
field $H_\mathrm{c2}$ of about 10 mT (see Ref.~\cite{ogata2026}). In
this forthcoming paper, comparing LaRh$_2$As$_2$ and CeRh$_2$As$_2$,
the authors suggest that the $4f$ electrons in the latter case not
only enhance the orbital limiting field but also play a role in the
development of unconventional superconductivity with SC multiphase.
Therefore, it can be concluded that our first scenario, which is
based on the exchange interaction between the $4f$ electrons of the
nearest neighbor Ce atoms, is more realistic.

\section{Conclusion}

The SOP symmetries of CeRh$_2$As$_2$ were  scrutinized on the basis
of Anderson functions for single pairs with taking into account
phase winding, time-reversal symmetry and the non-symmorphic
structure of the space group. It has been established that the
non-symmorphic structure of the space group of CeRh$_2$As$_2$
disrupts the direct relationship between the multiplicity and parity
of Cooper pairs at the high-symmetry points of the Brillouin zone.
This finding make possible to explain different spatial parities of
two superconducting phases within a single pairing interaction. It
was obtained that in axial symmetry  the nodes in vertical planes
may be removed by phase winding in the magnetic group symmetry
$4/mm^{\prime }m^{\prime}$, but the nodes in the basal plane are
unremovable. The spin-orbit coupling was considered in magnetic
symmetry, and it was obtained that nodeless SOP belongs to ICR
$E_{1u}^{\prime +}$  at generic point of the Brillouin zone in ESP
and OSP triplet cases. At the OSP $\rightarrow$ ESP transition, an
external magnetic field should destroy OSP pairs and stabilizes ESP
pairs. It was also obtained that triplet pairs at the X point are
even. Thus, the  SC1 and SC2 phases may be connected with even
triplet ESP pairs at the X point and odd triplet ESP pairs in the
$\Gamma-$X direction, respectively.  In this scenario spatial parity
changes without a changing of the parity with respect to permutation
of electronic coordinates and, consequently, without a changing of
the sign of the exchange integral between $4f$ orbitals of two
non-equivalent Ce atoms. This finding highlights the complexities
involved in understanding the behavior of Cooper pairs in
non-symmorphic systems.


\bibliography{refs}

\appendix

\setcounter{table}{0}
\renewcommand{\thetable}{A\arabic{table}}

\section{Calculation of two-electron states at symmetry planes, lines and points in a BZ}
\subsection{Basic equations}
One-electron states in a crystal with the symmetry group $G$ are
defined by the wavevector $k$, its symmetry group $H$ (little group)
and the index $\kappa $ of the IR $D^{\kappa }$ (small IR) of $H$
\cite{bc}. A left coset decomposition of the space group $G$ with
respect to its subgroup $H$ is written as:

\begin{equation}
G=\sum_{\sigma }s_{\sigma }H\text{.}  \label{lcda}
\end{equation}

The action of the left coset representatives $s_{\sigma }$ on the
wavevector $k$ results in a star $\left\{ k\right\} $ of this
wavevector. An IR of the space group $G$ is an induced
representation $D^{\kappa }\uparrow G$  defined by the following
formula \cite{bc}:

\begin{equation}
\left( D^{\kappa }\uparrow G\right)_{\sigma i\rho j} (g)=\left\{
\begin{array}{c}
D_{ij}^{\kappa }(s_{\sigma }^{-1}gs_{\rho }),\ \text{if }s_{\sigma
}^{-1}gs_{\rho
}\in H \\
0,\text{\ if }s_{\sigma }^{-1}gs_{\rho }\notin H%
\end{array}%
\right. \text{,}  \label{irspa}
\end{equation}

where we use Bradley's notation, with an up arrow for induction
\cite{bc}.

According to the Pauli exclusion principle, the state of two
equivalent fermions is antisymmetric with respect to permutation of
electrons' coordinates. Hence it follows that the spatial part of a
singlet (triplet) pair belongs to a symmetrized (antisymmetrized)
direct (Kronecker) square of the IR $\left( D^{\kappa }\uparrow
G\right) (g)$ of the space group. The decomposition of direct
squares can be easily done, making use of the Mackey-Bradley theorem
on symmetrized squares of induced representations
\cite{Mackey1953,Bradley1970,Yarzhemsky1992} as follows. The space
group $G$ is decomposed into double cosets with respect to the
wavevector group:

\begin{equation}
\mathbf{\ }G=\sum_{\delta }Hd_{\delta }H
\end{equation}

We consider Cooper pairs, i.e., pairs with zero total momentum $K$.
For any particular space group $G$ and a wavevector $k$, there are
different double cosets resulting in the zero total momentum of a
pair:

\begin{equation}
K=k+d_{\delta }k=b\text{ ,}
\end{equation}

where $b$ is a reciprocal lattice vector. For generic points inside
a BZ, $d_{\delta}$ is a space inversion, and we have:

\begin{equation}
K=k-k=0\text{ ,}
\end{equation}

On the faces of a BZ there are $k$ points for which:

\begin{equation}
K=k+k=b\text{ ,}
\end{equation}

In this case, the double coset is defined by the identity element
$E$, and two electrons with the same momenta form a pair. For
$d_{\delta}=E$ and symmorphic groups, symmetrization and
antisymmetrization of squares are performed over the group $H$
according to the standard formula \cite{ham}:

\begin{equation}
\chi^{\pm} \left( D^{\kappa} \times D^{\kappa}\right)(h) =\
\frac{\chi ^{2}\left( D^{\kappa
}(h)\right) \pm \chi \left( D^{\kappa }\left( h^{2}\right) \right) }{2}\text{%
.}  \label{sch1}
\end{equation}

In this case spatial parts of both singlet and triplet pairs are
even. In non-symmorphic space group for $k$ at the faces of a BZ IRs
$D^{\kappa}$ of $H$ are expressed through projective IRs
$\hat{D}^{\kappa}$ \cite{kov}. The above formula is extended to
projective IRs, by taking into account phase factors $\omega \left(
h,h\right)$. Thus, the characters on the wavevector group,
corresponding to symmetrized and antisymmetrized squares,
respectively, are written as \cite{ys21}:

\begin{equation}
\chi^{\pm} \left( \hat{D}^{\kappa}\times \hat{ D}^{\kappa}\right)(h) =\ \frac{\chi ^{2}\left( \hat{D}%
^{\kappa }(h)\right) \pm \omega \left( h,h\right) \chi \left( \hat{D}%
^{\kappa }\left( h^{2}\right) \right) }{2}\text{,}  \label{sch2}
\end{equation}

All possible IRs of the central extension $\tilde{G}$ corresponding
to singlet and triplet pairs are given by inducing these characters
into $\hat G$.

For each of the self-inverse double cosets, i.e., if:

\begin{equation}
Hd_{\alpha }H=Hd_{\alpha }^{-1}H\neq H
\end{equation}

we consider an element $a=dh_{1}=h_{2}d^{-1}$, where $h_{1},h_{2}\in
H$,  a subgroup:

\begin{equation}
\mathbf{\ }M_{\alpha }=aHa^{-1}\cap H\text{ }  \label{m}
\end{equation}

and its extension $\tilde{M}_{\alpha }$ by the element $a$:

\begin{equation}
\tilde{M}_{\alpha }=\mathbf{\ }M_{\alpha }+a\mathbf{\ }M_{\alpha
}\text{,}
\end{equation}

In the case of a self-inverse double coset, the characters of
representations $P_{\alpha }^{\kappa \pm }$ of $\tilde{M}_{\alpha}$,
correspond to symmetrized and antisymmetrized squares, respectively,
and are defined by the following two formulas:

\begin{equation}
\chi \left( P_{\alpha }^{\kappa \pm }(m)\right) =\chi \left( D^{\kappa }(m)%
\right) \chi \left( D^{\kappa }(a^{-1}ma)\right) \text{,  }m\in
M_{\alpha } \label{p1}
\end{equation}

\begin{equation}
\mathbf{\ }\chi \left( P_{\alpha }^{\kappa \pm }\left( am\right)
\right) \ =\pm \chi \left( D^{\kappa }(amam)\right) \text{, }m\in
M_{\alpha }\ \label{p2}
\end{equation}

In the case of Cooper pairs $d_{\alpha }=I$, and $P_{\alpha
}^{\kappa \pm}$ is a representation of the point group
$\tilde{M}_{\alpha}$. Induced representations $P_{\alpha }^{\kappa
\pm}\uparrow \hat{G}$ give possible symmetries of singlet (triplet)
pairs on the whole point group  $\hat{G}$. This reducible induced
representation can be easily decomposed into IRs $\Gamma ^{q}$ of
$\hat{G}$, making use of the Frobenius reciprocity theorem,
according to which the frequency $f$ of the appearance of $\Gamma
^{q}$ in the decomposition of the induced representation is given by
the following relation:

\begin{equation}
f(\Gamma ^{q}/P_{\alpha }^{\kappa \pm }\uparrow \hat{G})=\
\frac{1}{\left\vert \tilde{M}_{\alpha }\right\vert
}\sum_{\tilde{m}\in \tilde{M}_{\alpha}}\chi ^{\ast }\left( \Gamma
^{q}\left( \tilde{m}\right) \right) \chi \left( P_{\alpha }^{\kappa
\pm }(\tilde{m})\right) \text{,} \label{fr}
\end{equation}
where the vertical bars denote the number of elements in a group.

\subsection{Some data about groups $D_{4h}^{7}$ and $4/mm'm'$}

The IRs of the space groups of Kovalev \cite{kov} were used in
calculations of possible pairs' symmetries, and his digital
notations for the group elements $h_{i}$ were used in the examples.
The correspondence between the digital and standard notation is seen
in Table \ref{tab:Table A1}, where the elements of the space group
$D_{4h}^{7}$  are presented. The characters of IRs of the point
group $D_{4h}$ for reflections in planes are presented in Table
\ref{tab:Table A1b}. Negative characters correspond to nodal planes.
In the case of zero characters of two-dimensional IRs, the nodes on
planes depend on a unitary transformation. Real and complex forms of
two-dimensional IRs are presented in Table \ref{tab:Table A2}.

\begin{table*}[t!]
\caption{\label{tab:Table A1} Elements of the space group
$D_{4h}^{7}$ in standard and Kovalev \cite{kov} digital notations.
Subscripts $a$ and $b$ stand for axes $(110$) and $(-110)$
respectively, and also for the planes normal to these axes. }

\begin{tabular}{|l||l|l|l|l|l|l|l|l|}
\hline
& $E$, $C_{2z}$ & $C_{2x}$, $C_{2y}$ & $C_{2b}$, $C_{2a}$ & $C_{4z}$, $%
C_{4z}^{3}$ & $I$, $\sigma _{z}$ & $\sigma _{x}$, $\sigma _{y}$ &
$\sigma _{b}$, $\sigma _{a}$ & $S_{4z}$, $S_{4z}^{3}$ \\ \hline
Elements & $h_{1}$, $h_{4}$ & $h_{2}$, $h_{3}$ & $h_{13}$, $h_{16}$ & $%
h_{14}$, $h_{15}$ & $h_{25}$, $h_{28}$ & $h_{26}$, $h_{27}$ & $h_{37}$, $%
h_{40}$ & $h_{38}$, $h_{39}$ \\ \hline
$\tau $ & $\left( 000\right) $ & $\left( \frac{1}{2}\frac{1}{2}0\right) $ & $%
\left( \frac{1}{2}\frac{1}{2}0\right) $ & $\left( 000\right) $ &
$\left( \frac{1}{2}\frac{1}{2}0\right) $ & $\left( 000\right) $ &
$\left( 000\right)
$ & $\left( \frac{1}{2}\frac{1}{2}0\right) $\\
\hline
\end{tabular}

\end{table*}

\begin{table*}[t]
\caption{\label{tab:Table A1b} Characters of IRs of $D_{4h}$ for
reflections.  Negative characters correspond to group theoretical
nodal planes obtained by the Mackey-Bradley theorem (see Table
\ref{tab:table1}). In the case of zero characters, nodal planes
depend on a unitary transformation (see text and  Fig. \ref{fig2}).}

\begin{tabular}{|l|l|l|l|}
\hline & $\sigma _{x}$, $\sigma _{y}$ & $\sigma _{b}$, $\sigma _{a}$
& $\sigma _{z}$
\\ \hline
$A_{1g(u)}$ & $+1(-1)$ & $+1(-1)$ & $1(-1)$ \\ \hline $A_{2g(u)}$ &
$-1(+1)$ & $-1(+1)$ & $1(-1)$ \\ \hline $B_{1g(u)}$ & $+1(-1)$ &
$-1(+1)$ & $+1(-1)$ \\ \hline $B_{2g(u)}$ & $-1(+1)$ & $+1(-1)$ &
$+1(-1)$ \\ \hline $E_{g(u)}$ & 0 & $0$ & $-2(+2)$ \\ \hline
\end{tabular}

\end{table*}

\begin{table*}[t!]
\caption{\label{tab:Table A2} Matrices of two-dimensional IRs  of
the $D_{4h}$
group in real $E_{g(u)}$ and complex $\tilde{E}_{g(u)}$ forms. To obtain the matrices on the subgroup $\mathsf{C}%
_{4v}$ it is sufficient to multiply all matrices for pure rotations
by the
matrix of the left coset representative $h_{27}=\sigma _{y}$. Two extensions from $\mathsf
{C}_{4}$ to $\mathsf{C}_{4v}$, differing by the sign of the matrix for $h_{27}$
are possible. Note that on the subgroup $\mathsf{C}_{4v}$ the
matrices of $\tilde{E}_{g}$ and $\tilde{E}_{u}$ are chosen to be the
same. The matrices $\tilde{E}_{g}$ and $\tilde{E}_{u}$ may be easily
extended to $D_{4h}$ by multiplying them by inversion $I$ with the
signs $+$ or $-$, respectively. }

\begin{tabular}{|l|l|l|l|l|l|l|}
\hline
 IR & $h_{1}$ &  $h_{14}$ & $h_{4}$ & $h_{15}$ & $h_{27}$ & $h_{27}$\\ \hline
$E_{g(u)}$ & ${\left(
\begin{array}{cc}
1 & 0 \\
0 & 1%
\end{array}%
\right) }$ & ${\left(
\begin{array}{cc}
0 & -1 \\
1 & 0%
\end{array}%
\right) }$ & ${\left(
\begin{array}{cc}
-1 & 0 \\
0 & -1%
\end{array}%
\right) }$ & ${\left(
\begin{array}{cc}
0 & 1 \\
-1 & 0%
\end{array}%
\right) }$ & ${\left(
\begin{array}{cc}
0 & 1 \\
1 & 0%
\end{array}%
\right) }$ & ${\left(
\begin{array}{cc}
0 & -1 \\
-1 & 0%
\end{array}%
\right) }$  \\
\hline $\tilde{E}_{g(u)}$ & ${\left(
\begin{array}{cc}
1 & 0 \\
0 & 1%
\end{array}%
\right) }$ & ${\left(
\begin{array}{cc}
i & 0 \\
0 & -i%
\end{array}%
\right) }$ & ${\left(
\begin{array}{cc}
-1 & 0 \\
0 & -1%
\end{array}%
\right) }$ & ${\left(
\begin{array}{cc}
-i & 0 \\
0 & i%
\end{array}%
\right) }$ & ${\left(
\begin{array}{cc}
0 & 1 \\
1 & 0%
\end{array}%
\right) }$ & ${\left(
\begin{array}{cc}
0 & -1 \\
-1 & 0%
\end{array}%
\right) }$  \\
\hline
\end{tabular}

\end{table*}

The magnetic Shubnikov group with unitary subgroup $C_{4h}$    may
be written as:

\begin{equation}
4/mm^{\prime }m^{\prime }=C_{4h}+\vartheta \sigma _{y}\ C_{4h}\text{
}
\end{equation}

It is immediately verified  that the spatial parts of anti-unitary
elements $B$ are reflections in vertical planes and rotations around
horizontal axes by angles $\pi$, and hence $B^{2}=E$. Thus, the sum
in Herring's criterion \cite{bc} can be easily estimated:
\begin{equation}
\sum_{B\in \vartheta \sigma _{y}\mathsf{C}_{4h}}\chi \left(
B^{2}\right) =8 \label{herr}
\end{equation}

Hence, it follows from (\ref{herr}) that ICRs (irreducible
corepresentation)
belong to type (a), i.e., they are ordinary IRs for the unitary subgroup $%
C_{4h} $ and are extended to the non-unitary left coset with  plus
or minus sign:
\begin{equation}
\Delta ^{\pm }(R)=D(R),\;\Delta ^{\pm }(B)=\pm D(BA^{-1})N,\ N=1
\label{icr}
\end{equation}

\begin{table*}[t!]
\caption{\label{tab:Table A3}  ICRs of the magnetic group
$4/mm^{\prime }m^{\prime }$, and corresponding angular momentum
numbers $\bar{m}$. }

\begin{tabular}{|c|c|c|c|c|c|c|c|c|c|}
\hline
ICR & $E$ & $C_{4z}$ & $C_{2z}$ & $C_{4z}^{3}$ & $\vartheta \sigma _{y}$ & $%
\vartheta \sigma _{a}$ & $\vartheta \sigma _{x}$ & $\vartheta \sigma
_{b}$ & $\bar{m}$ \\ \hline
$A_{g(u)}^{+}$ & $1$ & $1$ & $1$ & $1$ & $+1$ & $+1$ & $+1$ & $+1$ & $0$, $%
\pm 4$ \\ \hline
$A_{g(u)}^{-}$ & $1$ & $1$ & $1$ & $1$ & $-1$ & $-1$ & $-1$ & $-1$ & $0$, $%
\pm 4$ \\ \hline $B_{g(u)}^{+}$ & $1$ & $-1$ & $1$ & $-1$ & $+1$ &
$-1$ & $+1$ & $-1$ & $\pm 2 $ \\ \hline $B_{g(u)}^{-}$ & $1$ & $-1$
& $1$ & $-1$ & $-1$ & $+1$ & $-1$ & $+1$ & $\pm 2 $ \\ \hline
$E_{1g(u)}^{\prime +}$ & $1$ & $i$ & $-1$ & $-i$ & $+1$ & $-i$ &
$-1$ & $i$ & $+1$ \\ \hline $E_{1g(u)}^{\prime -}$ & $1$ & $i$ &
$-1$ & $-i$ & $-1$ & $i$ & $+1$ & $-i$ & $+1$ \\ \hline
$E_{2g(u)}^{\prime +}$ & $1$ & $-i$ & $-1$ & $\ i$ & $+1$ & $i$ & $-1$ & $%
-i $ & $-1$ \\ \hline
$E_{2g(u)}^{\prime -}$ & $1$ & $-i$ & $-1$ & $\ i$ & $-1$ & $-i$ & $+1$ & $%
i $ & $-1$ \\
\hline
\end{tabular}

\end{table*}
 ICRs of the group $4/mm'm'$ are presented in Table \ref{tab:Table A3}.

\subsection{Symmetry planes in a BZ}

On the planes $k=b_{2}/2+\mu b_{1}+\nu b_{3}$ and $k=\mu b_{1}+\nu
b_{3}$  the little group $H$ consists of the identity element $E$
and the reflection $\sigma _{y}$. The factor system of the
projective representation of the group $H$ of the wavevector $k$ is
defined as follows:

\begin{equation}
\omega \left( \alpha ,\beta \right) =\exp (ib\cdot \tau _{\beta }),\
\ \label{wab1}
\end{equation}
\noindent where the reciprocal lattice vector $b$ is defined by the
relation:

\begin{equation}
\ \ k-\alpha ^{-1}k=b.  \label{wab2}
\end{equation}

It can be seen from (\ref{wab2}) that the first factor in formula (\ref%
{wab1}) differs from one, if $\alpha ^{-1}$ transforms the wave
vector $k$ into a vector that differs from it by a reciprocal
lattice vector $b$, and if the improper translation of the second
multiplier $\left\{ \beta |\tau _{\beta }\right\} $ has non-zero
component in a $b$ direction.

It can be immediately verified from the data of Table \ref{tab:Table
A1}, that on all faces of BZ IRs are not projective.

However, the space inversion  with an improper translation  appears
in formula (\ref{p1}), resulting in a phase factor. Indeed, using
multiplication rules for non-symmorphic group elements we obtain:

\begin{equation}
\left\{ I|\tau \right\} \left\{ \sigma _{y}|0\right\} \left\{ I|\tau
\right\} =\left\{ I|\tau \right\} \left\{ C_{2y}|\sigma _{y}\tau
\right\} =\left\{ \sigma _{y}|\tau -\sigma _{y}\tau \right\} \notag
\end{equation}
The translation of the resulting element is written as:
\begin{equation}
(\frac{1}{2}\frac{1}{2}0)-\sigma _{y}(\frac{1}{2}\frac{1}{2}0)=(\frac{1}{2}%
\frac{1}{2}0)-(\frac{1}{2}-\frac{1}{2}0)=(010)  \notag
\end{equation}

Since the  wavevectors on the plane $k=b_{2}/2+\mu b_{1}+\nu b_{3}$
have  a constant component $b_{2}/2$, the phase factor -1 appears.
In this case, formula (\ref{p2}) is written as:

\begin{equation}
\left\{C_{2y}|\tau \right\}\left\{C_{2y}|\tau \right\}  = \left\{
E|010\right\} \notag
\end{equation}

On the plane $k=b_{2}/2+\mu b_{1}+\nu b_{3}$  this factor also
equals to $-1$. On the plane $k=\mu b_{1}+\nu b_{3}$ located inside
the BZ, additional factors do not arise.

\begin{table*}[t!]
\caption{\label{tab:Table A5} Construction of Kronecker squares on
the plane $k=b_{2}/2+\mu b_{1}+\nu b_{3}$. }

\begin{tabular}{|c|c|c|c|c|c|c|}
\hline & $E$ & $\sigma _{y}$ & $I$ & $C_{2y}$ & IR$(\tilde{M})$ &
IR$(\hat{G})$ \\ \hline $P+$ & $1$ & $-1$ & $1$ & $-1$ & $B_{g}$ &
$A_{2g}+B_{2g}+E_{g}$ \\ \hline
$P^{-}$ & $1$ & $-1$ & $-1$ & $1$ & $A_{u}$ \  & $A_{1u}+B_{1u}+E_{u}$ \\
\hline
\end{tabular}

\end{table*}

\begin{table*}[t!]
\caption{\label{tab:Table A6} Construction of Kronecker squares on
the plane $k=\mu b_{1}+\nu b_{3}$. }

\begin{tabular}{|c|c|c|c|c|c|c|}
\hline & $E$ & $\sigma _{y}$ & $I$ & $C_{2y}$ & IR$(\tilde{M})$ &
IR$(\hat{G})$ \\ \hline $P^{+}$ & $1$ & $1$ & $1$ & $1$ & $A_{g}$ &
$A_{1g}+B_{1g}+E_{g}$ \\ \hline
$P^{-}$ & $1$ & $1$ & $-1$ & $-1$ & $A_{u}$ \  & $A_{2u}+B_{2u}+E_{u}$ \\
\hline
\end{tabular}

\end{table*}

\begin{table*}[t!]
\caption{\label{tab:Table A7} Construction of Kronecker squares on
the planes $k=\mu b_{1}+\nu b_{2}$ and $k=b_{3}/2+\mu b_{1}+\nu
b_{2}$. }

\begin{tabular}{|c|c|c|c|c|c|c|}
\hline & $E$ & $\sigma _{z}$ & $I$ & $C_{2z}$ & IR$(\tilde{M})$ &
IR$(\hat{G})$ \\ \hline $P^{+}$ & 1 & \ 1 & 1 & 1 & $A_{g}$ &
$A_{1g}+A_{2g}+B_{1g}+B_{2g}$ \\ \hline
$P^{-}$ & 1 & 1 & -1 & -1 & $A_{u}$ \  & $2E_{u}$ \\
\hline
\end{tabular}

\end{table*}

The data for the construction and the results for these planes are
presented in Tables \ref{tab:Table A5} and \ref{tab:Table A6}. It is
seen in these tables, that the non-symmorphic structure of the space
group changes the possible pair's symmetry on the lateral faces of
the BZ relative to parallel planes passing through the point
$\Gamma$. In particular, singlet pairs of $A_{1g}$ symmetry are
forbidden on the plane $k=b_{2}/2+\mu b_{1}+\nu b_{3}$, whereas on
the plane  $ k=\mu b_{1}+\nu b_{3}$, singlet pairs of $A_{1g}$
symmetry are possible. Since the improper translation $\left(
\frac{1}{2}\frac{1}{2}0\right) $  has no $z$ components, the results
for planes $k=\mu b_{1}+\nu b_{2}$ and $k=b_{3}/2+\mu b_{1}+\nu
b_{2}$ coincide. These results are presented in Table \ref{tab:Table
A7}.

\subsection{Symmetry points in a BZ}

\subsubsection{Pairing with equal $k$}

At points  M  and  A the wavevector group is a  $D_{4h}$ group.
Since the improper translation $\left(
\frac{1}{2}\frac{1}{2}0\right) $  has no component in z-direction,
the factor systems  and the results  for these points coincide. In
this case wavevector group coincides with the whole group $G$  and
symmetrization is performed by making use of formula (\ref{sch2}).
The data for the calculation and the results are presented in Table
\ref{tab:Table A8}. It is seen in this table, that at points $M$ and
$A$ singlet pairs (symmetrized squares) may be even and odd, but
triplet pairs are odd.

At points X and R space inversion belongs to the wavevector group
and two electrons with equal moments form a Cooper pair. In this
case symmetrization is performed by making use of formula
(\ref{sch2}). The data for calculations and the results for points X
and  R  are presented in Table \ref{tab:Table A9}. At these points,
singlet pairs may be even or odd, but for triplet pairs only $E_{g}$
symmetry is possible.

\begin{table*}[t!]
\caption{\label{tab:Table A8} The data for calculations and results
(last column) at points M and A } \
\begin{tabular}{|c|c|c|c|c|c|c|c|c|c|c|c|c|c|c|c|c|c|}
\hline
${\small h}$ & $h_{{\small 1}}$ & $h_{{\small 14}}$ & $h_{{\small 4}}$ & $h_{%
{\small 15}}$ & $h_{{\small 2}}$ & $h_{{\small 13}}$ & $h_{{\small 3}}$ & $%
h_{{\small 16}}$ & $h_{{\small 25}}$ & $h_{{\small 38}}$ &
$h_{{\small 28}}$
& $h_{{\small 39}}$ & $h_{{\small 26}}$ & $h_{{\small 37}}$ & $h_{{\small 27}%
}$ & $h_{{\small 40}}$ & {\small IR} \\
\hline ${\small h_{i}^{2}}$ & ${\small h_{1}}$ & ${\small h_{4}}$ &
${\small h_{1}}$ & ${\small h_{4} }$ & ${\small h_{1}}$ & ${\small
h_{1}}$ & ${\small h_{1}}$ & ${\small h_{1}}$ & ${\small h_{1} }$ &
${\small h_{4} }$ & ${\small h_{1} }$ & ${\small h_{4} }$ & ${\small h_{1}}$ & ${\small h_{1} }$ & $%
{\small h_{1}}$ & ${\small h_{1} }$ &  \\
\hline ${\small \omega _{ii} }$ & ${\small 1}$ & ${\small 1}$ &
${\small 1}$ & ${\small 1} $ & ${\small -1}$ & ${\small 1}$ &
${\small -1}$ & ${\small 1}$ & ${\small 1} $ & ${\small -1}$ &
${\small 1}$ & ${\small -1}$ & ${\small 1}$ & ${\small 1}
$ & ${\small 1}$ & ${\small 1}$ & \\
\hline ${\small t}_{1}$ & ${\small 2}$ & ${\small 0}$ & ${\small
-2}$ & ${\small 0}$ & ${\small 0}$ & ${\small -2}$ & ${\small 0}$ &
${\small 2}$ & ${\small 0}$ & ${\small 0}$ & ${\small 0}$ & ${\small
0}$ & ${\small 0}$ & ${\small 0}$ & ${\small 0}$ & ${\small 0}$ & \\
\hline ${\small t}_{2}$ & ${\small 2}$ & ${\small 0}$ & ${\small
-2}$ & ${\small 0}$ & ${\small 0}$ & ${\small 2}$ & ${\small 0}$ &
${\small -2}$ & ${\small 0}$ & ${\small 0}$ & ${\small 0}$ &
${\small 0}$ & ${\small 0}$ & ${\small 0}$ & ${\small 0}$ & ${\small
0}$ & \\ \hline ${\small t}_{3}$ & ${\small 2}$ & ${\small 0}$ &
${\small 2}$ & ${\small 0}$ & ${\small 0}$ & ${\small 0}$ & ${\small
0}$ & ${\small 0}$ & ${\small 0}$ &
${\small 0}$ & ${\small 0}$ & ${\small 0}$ & ${\small 0}$ & ${\small -2}$ & $%
{\small 0}$ & ${\small -2}$ & \\ \hline ${\small t}_{4}$ & ${\small
2}$ & ${\small 0}$ & ${\small 2}$ & ${\small 0}$ & ${\small 0}$ &
${\small 0}$ & ${\small 0}$ & ${\small 0}$ & ${\small 0}$ &
${\small 0}$ & ${\small 0}$ & ${\small 0}$ & ${\small 0}$ & ${\small 2}$ & $%
{\small 0}$ & ${\small 2}$ & \\
\hline
$\left[ t_{1,2}^{2}\right] $ & ${\small 3}$ & ${\small -1}$ & ${\small 3}$ & $%
{\small -1}$ & ${\small -1}$ & ${\small 3}$ & ${\small -1}$ &
${\small 3}$ &
${\small 1}$ & ${\small 1}$ & ${\small 1}$ & ${\small 1}$ & ${\small 1}$ & $%
{\small 1}$ & ${\small 1}$ & ${\small 1}$ & ${\small A_{1g}+B_{2g}+B_{2u}}$ \\
\hline
$\left\{ t_{1,2}^{2}\right\} $ & ${\small 1}$ & ${\small 1}$ & ${\small 1}$ & $%
{\small 1}$ & ${\small 1}$ & ${\small 1}$ & ${\small 1}$ & ${\small 1}$ & $%
{\small -1}$ & ${\small -1}$ & ${\small -1}$ & ${\small -1}$ &
${\small -1}$
& ${\small -1}$ & ${\small -1}$ & ${\small -1}$ & ${\small A_{1u}}$ \\
\hline
$\left[ t_{3,4}^{2}\right] $ & ${\small 3}$ & ${\small 1}$ & ${\small 3}$ & $%
{\small 1}$ & ${\small -1}$ & ${\small 1}$ & ${\small -1}$ & ${\small 1}$ & $%
{\small 1}$ & ${\small -1}$ & ${\small 1}$ & ${\small -1}$ & ${\small 1}$ & $%
{\small 3}$ & ${\small 1}$ & ${\small 3}$ & ${\small A_{1g}+B_{2g}+A_{2u}}$ \\
\hline $\left\{ t_{3,4}^{2}\right\} $ & ${\small 1}$ & ${\small -1}$
& ${\small 1}$ & ${\small -1}$ & ${\small 1}$ & ${\small -1}$ &
${\small 1}$ & ${\small -1}$ & ${\small -1}$ & ${\small 1}$ &
${\small -1}$ & ${\small 1}$ & ${\small -1}$
& ${\small 1}$ & ${\small -1}$ & ${\small 1}$ & ${\small B_{1u}}$ \\
\hline
\end{tabular}

\end{table*}

\begin{table*}[t!]
\caption{\label{tab:Table A9} The data for calculation and  results
(last column) for points X ($k=\frac{b_{2}}{2}$ \cite {kov}) and R
($k=\frac{b_{2}}{2}+\frac{ b_{3}}{2}$ \cite {kov}). }

\begin{tabular}{|c|c|c|c|c|c|c|c|c|c|}
\hline
IR $X$, $R$ & $h_{1}$ & $h_{3}$ & $h_{2}$ & $h_{4}$ & $h_{25}$ & $h_{27}$ & $%
h_{26}$ & $h_{28}$ & IR \\ \hline $t_{1}$ & $2$ & $0$ & $0$ & $0$ &
$0$ & $0$ & $2$ & $0$ &  \\ \hline $t_{2}$ & $2$ & $0$ & $0$ & $0$ &
$0$ & $0$ & $-2$ & $0$ &  \\ \hline $\omega_{ii} $ & $1$ & $1$ &
$-1$ & $1$ & $-1$ & $1$ & $1$ & $1$ &  \\ \hline
$\left[ t_{1,2}^{2}\right] $ & $3$ & $1$ & $-1$ & $1$ & $-1$ & $1$ & $3$ & $%
1 $ & $A_{1g}+A_{2u}\ +B_{1g}+B_{2u}+E_{u}$ \\ \hline $\left\{
t_{1,2}^{2}\right\} $ & $1$ & $-1$ & $1$ & $-1$ & $1$ & $-1$ & $1$
& $-1$ & $E_{g}$\\
\hline
\end{tabular}

\end{table*}

\subsubsection{Pair density waves at point X}

For the double coset, defined by the  reflection in a diagonal
plane, we have the relation:

\begin{equation}
k_{X}+\sigma_{b}k_{X}=K_{M}  \notag
\end{equation}

Such two-electron states do not have complete translational symmetry
and can be related to pair density waves. The data for the
calculation of pair density waves at point X  and the results are
presented in Table \ref{tab:Table A10}.  We need to estimate in the
product additional phase factors for the elements with non-zero
characters  of IRs at point X. It so happens that the transformation
by element $h_{37}$ permutes elements $h_{26}$ and $h_{27}$. The
characters of the projective representations $t_{1,2}$ at point X
are nonzero only for $h_{26}$ and equal to zero for $h_{27}$ (see
Table \ref{tab:Table A9}), so there is no need to calculate
additional factors for the terms
$\hat{D}(d^{-1}_{\delta}h_{i}d_{\delta}) \hat{D}(h_{i})$. Going to
the squares of left coset element in Mackey-Bradley theorem (see formula
(\ref{p2}) we ensure that, since $h_{37}$ and $h_{40}$ appear
without improper translation, the phase factors for the terms
$h_{37}^{2}$ and $h_{40}^{2}$  equal to  unity.

Bearing in mind that:

\begin{equation}
h_{13}\cdot \left( \frac{1}{2}\frac{1}{2}0\right) =\left(
-\frac{1}{2}- \frac{1}{2}0\right) \text{,} \notag
\end{equation}
and that:
\begin{equation}
h_{16}\cdot \left( \frac{1}{2}\frac{1}{2}0\right) =\left(
\frac{1}{2} \frac{1}{2}0\right)   \notag
\end{equation}

we can easily estimate the products:

$\left\{ h_{13}|\frac{1}{2}\frac{1}{2}0\right\} \cdot \left\{
h_{13}|\frac{ 1}{2}\frac{1}{2}0\right\} =\left\{ h_{1}|000\right\} $

$\left\{ h_{16}|\frac{1}{2}\frac{1}{2}0\right\} \cdot \left\{ h_{16}|\frac{%
1}{2}\frac{1}{2}0\right\} =\left\{ h_{1}|111\right\} $

Hence it follows that the phase factors $-1$ and $1$ appear for the
terms  $h_{16}^{2}$ and  $h_{13}^{2}$ respectively.

\begin{table*}[t!]
\caption{\label{tab:Table A10} The data for the calculation of pair
density waves     ($d_{\delta}=\sigma_{b}=h_{37}$) at point X.
$K_{\delta}=K_{M}$}.

\begin{tabular}{|c|c|c|c|c|c|c|c|c|c|c|}
\hline
$h_{i}$ & $h_{1}$ & $h_{3}$ & $h_{2}$ & $h_{4}$ & $h_{25}$ & $h_{27}$ & $%
h_{26}$ & $h_{28}$ & formula & IR(M) \\ \hline
$d^{-1}_{\delta}h_{i}d_{\delta}$ & $h_{1}$ & $h_{2}$ & $h_{3}$ &
$h_{4}$ & $h_{25}$ & $h_{26}$ & $h_{27}$ & $h_{28}$ &  &  \\ \hline
$\chi ^{\pm }(h_{i})$ & $4$ & $0$ & $0$ & $0$ & $0$ & $0$ & $0$ & $0$ & (\ref%
{p1}) &  \\ \hline $d_{\delta}h_{i}$ & $h_{37}$ & $h_{38}$ &
$h_{39}$ & $h_{40}$ & $h_{13}$ & $h_{14}$ & $h_{15}$ & $h_{16}$ &  &
\\ \hline $d_{\delta}h_{i}d_{\delta}h_{i}$ & $h_{1}$ & $h_{4}$ &
$h_{4}$ & $h_{1}$ & $h_{1}$ & $h_{4}$ & $h_{4}$ & $h_{1}$ &  &  \\
\hline
$\chi ^{+}(d_{\delta}h_{i})$ & $2$ & $0$ & $0$ & $2$ & $2$ & $0$ & $0$ & $-2$ & (\ref%
{p2}) & $t_{2}+t_{4}$ \\ \hline
$\chi ^{-}(d_{\delta}h_{i})$ & $-2$ & $0$ & $0$ & $-2$ & $-2$ & $0$ & $0$ & $2$ & (%
\ref{p2}) & $t_{1}+t_{3}$ \\
\hline
\end{tabular}

\end{table*}

These two-electron states at  the M have point group symmetry
$D_{4h}$, but since their total momentum is
$(\frac{b_{1}}{2}\frac{b_{2}}{2}0)$, they are totally symmetric with
respect to half of lattice translations only.

\end{document}